\documentclass{aa}
\usepackage{graphicx}
\usepackage{txfonts}
\usepackage{natbib}

\newlength{\dlugskr}

\def\figref#1{Fig.\,\ref{#1}}

\def\normal{{\em normal}}

\def\DCcr{DC{\em{$_{cr}$}}}
\def\DCacr{DC{\em{$_{a+cr}$}}}
\def\OCacr{OC{\em{$_{a+cr}$}}}

\def\um{$\mu$m}
\def\Msun{M$_{\sun}$}

\begin{document}
\title{New groups of planetary nebulae with peculiar
       dust chemistry towards the Galactic bulge\thanks{Based on 
observations made with the Spitzer Space Telescope, which is operated by the 
Jet Propulsion Laboratory, California Institute of Technology, under NASA 
contract 1407.} }
\titlerunning{New PNe with peculiar dust chemistry}
   \author{S.K. G\'orny\inst{1} \and 
           J. V. Perea-Calder\'on\inst{2} \and 
           D. A. Garc\'{\i}a-Hern\'andez\inst{3} \and 
           P. Garc\'{\i}a-Lario\inst{4} \and 
           R. Szczerba\inst{1}
}
	  \authorrunning{G\'orny et al.}
  \offprints{S.K. G\'orny}

  \institute{N. Copernicus Astronomical Center, Rabia\'nska 8, 87-100 Toru\'n, Poland
\email{skg@ncac.torun.pl}
   \and European Space Astronomy Centre, INSA S.A.  
     P.O. Box 50727. E$-$28080 Madrid. Spain 
  \and Instituto de Astrof\'{\i}sica de Canarias, C/ Via L\'actea s/n,
     38200 La Laguna, Spain  
  \and Herschel Science Centre. European Space Astronomy Centre, Research and
       Scientific Support Department of ESA. Villafranca del Castillo, 
       P.O. Box $-$ Apdo.50727. E$-$28080 Madrid. Spain.
}

   \date{Received 29 July 2009; accepted 15 January 2010}
%
%
  \abstract
  {}
  {We investigate Galactic bulge planetary nebulae without emission-line
  central stars for which peculiar infrared spectra have been obtained
  with the Spitzer Space Telescope, including the simultaneous signs of oxygen and carbon based
  dust. Three separate sub-groups can be defined characterized by the
  different chemical composition of the dust and the presence of crystalline
  and amorphous silicates.}
  {We use literature data to analyze the different nebular properties and
  deduce both the evolutionary status and the origin of these three groups.
  In particular, we check whether there are signs of evolutionary links
  between dual-dust chemistry planetary nebulae without detected
  emission-line central stars and those with emission-line stars.}
  {Our primary finding is that the classification based on the dust
  properties is reflected in the more general properties of these
  planetary nebulae.  However, some observed properties are difficult to
  relate to the common view of planetary nebulae. In particular, it is
  challenging to interpret the peculiar gas chemical composition of many
  analyzed objects in the standard picture of the evolution of planetary 
  nebulae
  progenitors. We confirm that the dual-dust chemistry phenomenon is not
  limited to planetary nebulae with emission-line central stars.}
  {Our results clearly indicate that there is no unique road to the
  formation of planetary nebulae even in a homogeneous environment such as
  the Galactic bulge.  The evolution of a single asymptotic giant branch
  star may lead to the formation of different types of planetary nebulae.
  In addition, the evolution in a close binary system should
  sometimes also be considered.}

%

\maketitle

\section{Introduction}

After the completion of hydrogen and helium burning in their cores, low- to
intermediate-mass stars (0.8 $\leq$ M $\leq$ 8 \Msun) evolve towards the
asymptotic giant branch \citep[AGB; e.g.,][]{Herwig2005} and then pass
through the planetary nebula (PN, plural PNe) phase before ending their
lives as white dwarfs. At the tip of the AGB phase, these stars experience
strong mass loss that efficiently enriches the surrounding interstellar
medium with huge amounts of gas and dust. Stars leave the AGB when the
strong mass loss stops and then the future central star (CS, plural CSs)
rapidly evolves towards hotter effective temperatures in the
Hertzsprung-Russell diagram. Thus, when the ionization of the ejected gas
takes place, a new PN is formed. However, in most cases the total amount of
ionized gas is very small compared to the total mass previously ejected. An
important fraction of this material remains neutral in the form of dust
grains, molecules, or atoms, which can be easily detected in the infrared
domain. Thanks to the analysis of the features observed by the Infrared
Space Observatory (ISO) in the spectra of PNe, it was possible to confirm
the presence of large amounts of dust grains around PNe as well as the
dominant dust chemistry (C-rich versus O-rich). Features at 3.3,
6.2, "7.7", 8.6, and 11.3\um\ attributed to polycyclic aromatic
hydrocarbons (PAHs) are common in C-rich PNe
\citep[e.g.,][and references therein]{Garcia-Lario1999} while strong features
attributed to crystalline silicates (e.g., those centered on 23.5, 27.5 and
33.8 \um) are usually found in O-rich PNe
\citep[e.g.,][]{Molster2002}.

A few Galactic disk PNe exhibited a remarkable dual-dust (C-rich and O-rich)
chemistry showing both PAH and crystalline silicate features in ISO
spectra \citep{Waters1998a, Waters1998b, Cohen1999, Cohen2002}. The fact
that this was an infrequent phenomenon may be due to the instruments used,
which in many cases may have been unable to detect crystalline silicates.
For example, the Spitzer Space Telescope \citep[Spitzer,][]{Werner2004},
detected crystalline silicates in 10 post-AGB sources
\citep{Cerrigone2009} while after completion of the ISO mission only 2 such
sources were known \citep{Szczerba2003}.

The mixed-chemistry PNe discovered by ISO pertain to the class of objects
with C-rich Wolf-Rayet type nuclei (the so-called [WR] PNe), which usually
show a lack of hydrogen in their atmospheres. These atmospheres are instead
mostly composed of helium, carbon, and oxygen and the CSs show intense
mass-loss \citep[e.g.,][]{Crowther2008}.

The evolution of an AGB star with a stellar or substellar companion that
undergoes the common-envelope phase is another possible way of creating a
PN. Some authors argue that a companion object is
often mandatory for a planetary nebula to be created
\citep[see in][and references therein]{DeMarco2009}. We noted that
some hypotheses compiled to explain the simultaneous presence of carbon and
oxygen dust also require a binary system.

The Galactic bulge is characterized by an old population of mostly low-mass
stars (\citet{Zoccali2003}, but see also \citet{Uttenthaler2007} and
references therein).  It is also known that the abundances of PNe in the
Galactic bulge (GBPNe) differ from those located in the Galactic disk
as they have higher metallicities and lower C/O ratios
\citep[e.g.,][]{WangLiu2007}.  The differences in metallicity
seems to play a dominant role in the chemical evolution of low- to
intermediate-mass stars (e.g., \citet{Garcia-Hernandez2007};
\citet{Stanghellini2007}; \citet{Chiappini2009}). Thus, studding the Galactic
bulge enables us to investigate the stellar evolution of low- and
intermediate-mass stars in higher metallicity environments and at the same
time an insight into the chemical evolution and formation of our Milky Way.

\citet{Gutenkunst2008} analyzed Spitzer spectra acquired using the Infrared Spectrograph
\citep[IRS,][]{Houck2004} of 11 PNe towards the Galactic bulge and inferred
dual-dust chemistry in 6 of them.  They suggested that the high percentage
of dual-dust chemistry sources is unsurprising because the
fraction of [WR] PNe is significantly higher in the bulge than in the
Galactic disk. However, as we checked, only one of their dual-dust chemistry
sources have the wind characteristics of [WR] type CS and the
higher proportion of genuine [WR] PNe in the Galactic bulge is not confirmed
\citep{Gorny2009}. \citet{P-C2009} found that dual-dust chemistry is truly widespread
among GBPNe.  They analyzed a
larger sample of 26 GBPNe observed with Spitzer/IRS among which 21 exhibit
dual-dust chemistry. \citet{P-C2009} observations shown that the
simultaneous presence of oxygen and carbon-rich dust features in the
infrared spectra of [WR] PNe is not restricted to objects with
late/cool [WC] class stars. In addition, dual-dust chemistry was seen in all
observed PNe with WEL stars ({\em weak emission-line} stars,
\citealp{Tylenda1993}) as well as members of recently discovered VL group
(low ionization PNe around stars with {\em very late} [WC\,11]-like spectra,
\citealp{Gorny2009}). Surprisingly, \citet{P-C2009} found
dual-dust chemistry also in some PNe without detected emission-line CSs.

Another interesting property of the PNe observed by \cite{P-C2009} was the
amorphous silicate emission at 10\,$\mu$m, which was detected in four
dual-dust chemistry GBPNe and in most of the O-rich PNe that they observed.
Note that before Spitzer there was known only one such PN, namely SwSt\,1
\citep[e.g.,][]{Szczerba2001}, belonging to the [WR]~PNe. In contrast, the
10\um\ feature objects found by \cite{P-C2009}, neither belong
to this group nor exhibit stellar emission lines.

In this work, we analyze the multiple properties of PNe without
emission-line CSs that are found to have peculiar infrared spectra acquired
with Spitzer/IRS. The paper is organized as follows. In Sect.~2 and
3 we describe our working sample and the main properties of their
infrared spectra, respectively. We analyze the nebular properties and
evolutionary status of these PNe in Sect.~4. Finally, in Sect.~5 we
discuss the results obtained. Our concluding remarks are given in Sect.~6.

\begin{table} \caption{
  List of analyzed PNe. The Galactic bulge (b) or disk (d) association of
  the object is marked in Col.~3. The reference to original observers is
  given in Col.~4. Objects are divided into three groups according to
  their infrared properties as marked in Cols.~5--7 (see description in
  Sect.~3).}
\begin{tabular}{
               l @{\hspace{0.40cm}}
               l @{\hspace{0.40cm}}
               c @{\hspace{0.40cm}}
               c @{\hspace{0.60cm}}
               c @{\hspace{0.20cm}} c @{\hspace{0.20cm}} c}
\hline
\hline
   PN G     & name     &  pop.  & Ref. & \multicolumn{3}{c}{IR spectra} \\
            &          &        &      & PAH & SiO cr. & SiO am.        \\
\hline
 \DCcr:     &          &        &      &     &         &                \\
 000.1+04.3 & H 1-16   &    b   &  P   &  +  &    +    &    -           \\
 007.2+01.8 & Hb 6     &    d   &  P   &  +  &    +    &    -           \\
 354.2+04.3 & M 2-10   &    b   &  G   &  +  &    +    &    -           \\
 358.7-05.2 & H 1-50   &    b   &  P   &  +  &    +    &    -           \\
\vspace{0.15cm}
 358.9+03.2 & H 1-20   &    b   &  G   &  +  &    +    &    -           \\
 \DCacr:    &          &        &      &     &         &                \\
 354.5+03.3 & Th 3-4   &    b   &  P   &  +  &    +    &    +           \\
 356.9+04.4 & M 3-38   &    b   &  P   &  +  &    +    &    +           \\
 358.2+04.2 & M 3-8    &    b   &  P   &  +  &    +    &    +           \\
\vspace{0.15cm}
 359.7-02.6 & H 1-40   &    b   &  P   &  +  &    +    &    +           \\
 \OCacr:    &          &        &      &     &         &                \\
 002.2-02.7 & M 2-23   &    b   &  P   &  -  &    +    &    +           \\
 008.2+06.8 & He 2-260 &    d   &  P   &  -  &    +    &    +           \\
 355.6-02.7 & H 1-32   &    b   &  P   &  -  &    +    &    +           \\
 355.7-03.5 & H 1-35   &    b   &  P   &  -  &    +    &    +           \\
\hline
\hline
\end{tabular}
Ref: G -- Gutenkunst et~al. (2008); P -- Perea-Calder{\'o}n et~al. (2009)
\end{table}

\section{Sample selection}

We analyze the various properties of PNe without emission-line CSs observed
with Spitzer/IRS and exhibiting signs of dual-dust chemistry and/or 10\um\
emission band of amorphous silicates. These PNe are listed in Table~1 along
with a short description of their infrared features.  Most of them
were observed by \cite{P-C2009}, while two come from \cite{Gutenkunst2008}.

All of the selected PNe were included in the analyses of \cite{Gorny2009}
but none were attributed to either WEL, VL, or [WR]-type
groups of GBPNe. The absence of emission-line CSs was perceived in
these PNe by studying high quality optical spectra acquired at 2 and 4-meter
telescopes by
\citet{Cuisinier2000}, \citet{Escudero2001}, \citet{Escudero2004},
\citet{Gorny2004}, and new observations in \cite{Gorny2009}, or by checking
the list of the observed lines in
\citet{WangLiu2007}. The searched stellar emission lines were the same as in
\cite{Gorny2004}. In none of the thirteen PNe presented in Table~1 did we
notice any characteristics of emission-line CSs during
additional direct re-inspection of the spectra.

The objects for which the CS was of unknown type are not
considered until their spectral classification is established. The PNe with
dual-dust chemistry but unknown spectral type of the CS are
H\,2-20 and M\,2-5 observed by Gutenkunst et~al. (2008)\footnote{The
remaining dual-dust chemistry PNe on their list belong to [WR] type
(M\,2-31) and WEL PNe (H\,2-11).} and H\,1-62 by \cite{P-C2009}. The
spectral type of the CS is also unknown for three O-rich PNe
with the detected 10\um\ emission feature, i.e., He\,3-1357, Cn\,1-3 and
IC\,4732 listed by \cite{P-C2009}.

Eleven of the PNe listed in Table~1 have a high probability of physically
belonging to the Milky Way bulge because they satisfy the standard criteria
(Stasi\'nska \& Tylenda 1994), namely: a) they are located within 10 degrees
of the center of the Galaxy, b) have diameters smaller than 20\arcsec, and
c) the known radio fluxes at 5GHz are lower than 100\,mJy. 
The two remaining objects are either too bright in the radio domain (Hb 6)
or marginally outside the 10 degree angular distance (He 2-260) and
therefore regarded as possible Galactic disk members.

The 13 objects collected in Table~1 are compared in this paper with the
remainder of the total of 180 GBPNe analyzed in \cite{Gorny2009}. Among this
large reference group, there are 119 PNe without emission-line CSs (which we
refer to as \normal\ PNe throughout this paper), 25 WEL, 14 VL, and 9 [WR]
PNe. Some of the PNe from the latter groups were also observed with
Spitzer/IRS by either the Perea-Calder\'on or Gutenkunst teams. This
includes 5 WEL, 4 [WR], and 3 VL~PNe.

In this paper, we compare the objects investigated here to both
\normal\ PNe and PNe with emission-line CSs since it can
not be excluded that the latter, in particular [WR] PNe\footnote{
 In the Galactic bulge, only [WR] PNe with CSs of intermediate [WC]
 spectral classes are known, in contrast to the situation in the Galactic disk
 \citep[see e.g.,][]{Gorny2001}},
have at times weaker stellar winds and emission lines from CSs are not
observed. It has to be checked if this is not the case of some of the PNe
from Table~1. On the other hand, the stellar emission lines of WEL and VL
PNe are not very strong and may escape unnoticed in spectra of lower signal
to noise. Therefore, to confirm that indeed dual-dust chemistry phenomenon
also occurs in PNe without emission-line CSs we compare their properties
with those of [WR], WEL, and VL PNe.

Among the largest group of 119 \normal\ bulge PNe that we use as a reference
sample only four have been observed with Spitzer/IRS. This is a clear result
of the selection effect since the targets for observations were chosen by
\citet{P-C2009} based on high quality IRAS fluxes at 12, 25, and 60\,{\um}.
In particular, the requirement of reliable measurements at 12\,{\um} is
fulfilled by only 15 \normal\ PNe of 64 with IRAS data available. This means
that no prominent emission features around 12\,{\um} should be expected in
\normal\ PNe. Even in the case of the four objects observed by Spitzer, only
crystalline silicates are present in their spectra and both PAH and
amorphous silicates seem to be absent\footnote{Among these PNe, object
M\,3-13 has exceptional Spitzer spectra with unidentified broad emission
features at shorter wavelengths, see Fig.B4 of \cite{P-C2009}}.
Unfortunately, the existing optical spectra of these four PNe are at the
same time of low quality implying e.g., that the nature of their CSs is 
uncertain and many parameters cannot be reliably determined.

We add to the discussion the analysis of two PNe from the Galactic disk
population. One is the [WR]-type SwSt\,1 mentioned already in the
introduction and the other is IC\,4776 that has a WEL CS and
was observed by \cite{P-C2009}. Both these objects exhibit the dual-dust
chemistry signatures typical of PNe with emission-line CSs but additionally
also the 10\um\ amorphous silicate feature.

\begin{figure*}
\resizebox{0.95\hsize}{!}{\includegraphics{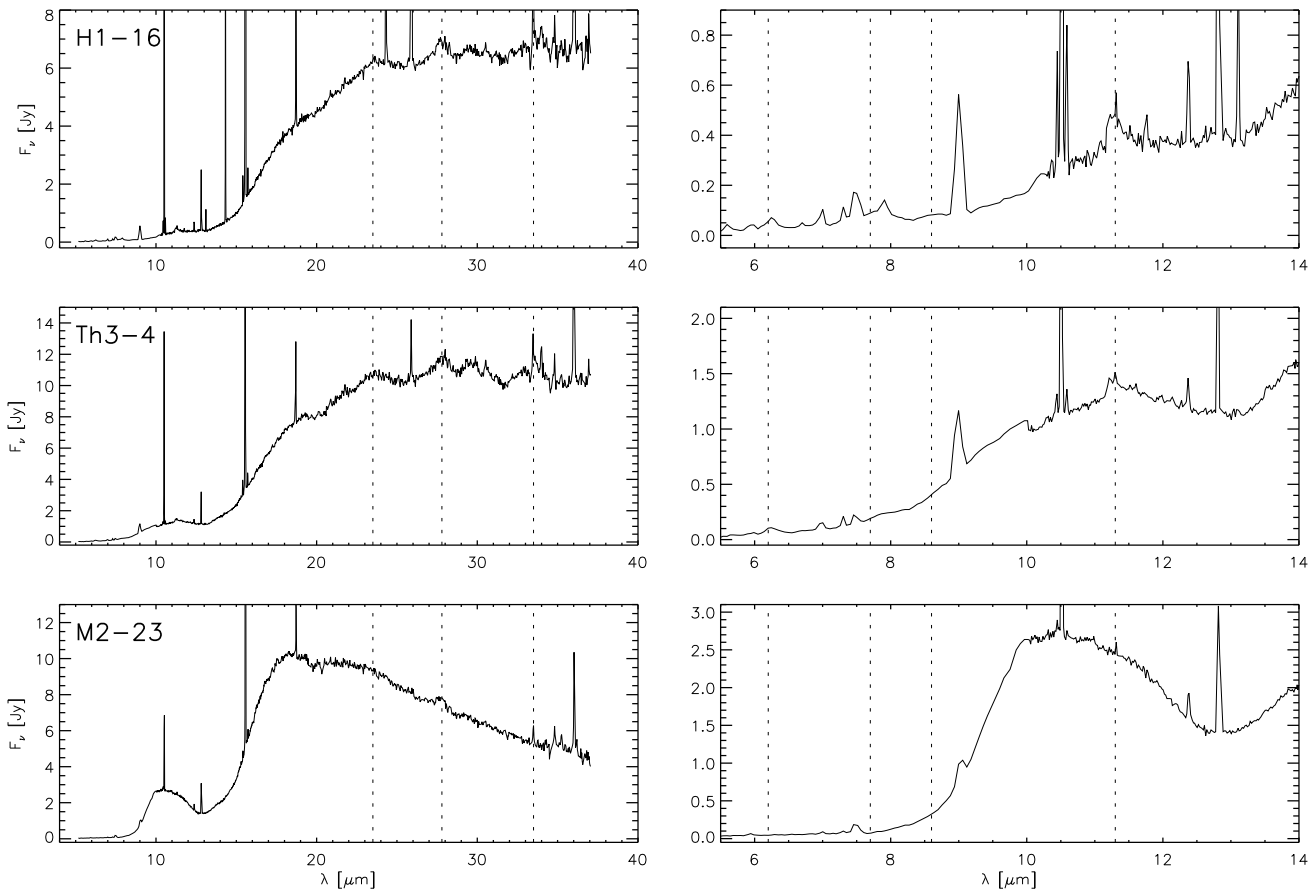}}
\caption[]{
  Examples of Spitzer/IRS spectra of PNe with three different types of dust
  composition: H\,1-16 of \DCcr\ type at the top, Th\,3-4 of \DCacr\ in the
  middle, and M\,2-23 of \OCacr\ type at the bottom. Dotted vertical lines
  indicate the positions of crystalline silicate emission features at 23.5,
  27.5, and 33.8 \um\ (left panels) and typical PAH features at 6.2, "7.7",
  8.6, and 11.3 \um\ (right panels). The 10\um\ amorphous silicate feature is
  seen in the middle and bottom spectra.
}
\label{IR_spec}
\end{figure*}

\section{Infrared spectra}

The infrared spectra of the GBPNe without emission-line CSs selected for
this work were acquired by \cite{P-C2009} and \citet{Gutenkunst2008}. The
description of the Spitzer/IRS data reduction process is not repeated here
and we refer the reader to these two references for details.

After a short inspection of the Spitzer/IRS spectra, we find that among PNe
without emission-line CSs, there are three clearly different types of
objects. The examples are presented in \figref{IR_spec}. It can be seen that
H\,1-16 and Th\,3-4 PNe have mixed chemistry where both PAHs and
crystalline silicates features are detected. For Th\,3-4 we also detect the
amorphous silicate emission feature at 10 \um. On the other hand, M\,2-23 is
clearly O-rich (there is a lack of carbon-based dust features such as PAHs),
exhibiting the amorphous silicate feature at 10 \um\ and weak crystalline
silicate features at longer wavelengths. Finally, we note that
M\,2-23 contains a circumstellar dust envelope, which is hotter than in
H\,1-16 and Th\,3-4 as can be inferred from the shape of continuum emission
in the spectrum.

Based on the examples from \figref{IR_spec}, we can describe the spectra of
all PNe analyzed in this work and divide them into different subgroups
(Table~1). Each group has some similarity with one of the other groups and
clear dissimilarity with another. The first group of 5 PNe listed at the top
of Table~1 is characterized by the simultaneous presence of both
carbon-based dust (PAH features at 6.2, "7.7", 8.6, and 11.3\,\um) and
oxygen-based dust (crystalline silicate features at 23.5, 27.5, and
33.8\,\um). We refer to this group as \DCcr\ ({\em dual-dust chemistry} with
silicates only in {\em crystalline} form) subsample throughout the rest of
this paper. The second group in Table~1 has 4 members and is also
characterized by the simultaneous presence of oxygen and carbon dust but at
the same time there are also signs of amorphous silicate features at about
10\um. We therefore refer to this group as \DCacr\ ({\em dual-dust
chemistry} with silicates in {\em amorphous} and {\em crystalline} form).
Finally, 4 PNe in Table~1 exhibit only oxygen dust features, which however
include the uncommon 10\um\ amorphous silicate feature. This last group
is called \OCacr\ ({\em oxygen-dust chemistry} with {\em amorphous} and
{\em crystalline} forms) subsample.

Detailed comparison and analysis of dust features observed with Spitzer/IRS
in GBPNe will be presented in the forthcoming paper
(Szczerba et al. {\em in prep.})

\begin{table*}
\begin{minipage}[t]{1.95\columnwidth}
\caption{Observational data for analyzed planetary nebulae. The reported
 quantities are: logarithmic extinction at H$\beta$ (Col.\,2); ionization
 parameters O$^{++}$/(O$^{+}$+O$^{++}$) and He$^{++}$/(He$^{+}$+He$^{++}$)
 (Col.\,3 and 4); electron density from [S\,II] $\lambda$6717/6731 line
 ratio (Col.\,5); nebular diameter (Col.\,6); nebular expansion velocity
 (Col.\,7); kinematic age (Col.\,8); H$\beta$ surface brightness (Col.\,9);
 stellar temperature from Zanstra hydrogen method (Col.\,10); mass of ionized
 nebular gas (Col.\,11) and mass of dust (Col.\,12).
}
\centering
\renewcommand{\footnoterule}{}
\begin{tabular}{
               l
               c @{\hspace{0.20cm}}
               c 
               c c l l c c c c c c c }
\hline
\hline
  name    & C$_{opt}$ & O$^{++}$/O  & He$^{++}$/He & log\,n$_e$ & $\theta$ & V$_{exp}$    & {\em{t}}$_{kin}$ & log S$_{H\beta}$ & log T$_{Zan}$ & M$_{gas}$ & M$_{dust}$ \\
\hline
            &         &             &              & [cm$^{-3}$] & [\arcsec]& [km/s]       & [year]    & [erg\,cm$^{-2}$s$^{-1}$sr$^{-1}$] & [K]& [M$_\odot$] & [10$^{-3}$\,M$_\odot$ ] \\
\hline
  H 1-16    & 2.50    & 0.880       & 0.097        & 3.77        & 2.0      & 20 $^{R}$    & 2000 $^r$ & -0.83            &      -           & 0.152     & 0.38       \\
  Hb 6      & 2.10    & 0.901       & 0.115        & 3.53        & 6.0      & 20 $^{RRA}$  &~~500 $^s$ & -0.95            & $>$ 4.48         & {\em 0.2} & {\em 1.20} \\
  M 2-10    & 1.32    & 0.625       & -            & 3.10        & 4.2      &              &           & -2.00            &      -           & 0.208     & 1.00       \\
  H 1-50    & 0.68    & 0.959       & 0.093        & 3.69        & 1.4      & 23 $^{GZ,R}$ & 1300      & -0.62            &     4.77         & 0.148     & 0.18       \\
\vspace{0.1cm}
  H 1-20    & 2.31    & 0.896       & 0.013        & 3.57        & 3.3      & 18 $^{R}$    & 3800 $^r$ & -1.30            &      -           & 0.223     & 0.68       \\
  Th 3-4    & 2.83    & 0.946       & 0.078        & 4.20        & 1.5      &              &           & -0.60            &      -           & 0.055     & 0.49       \\
  M 3-38    & 2.06    & 0.968       & 0.238        & 3.49        & 1.8      & 15 $^{R}$    & 2400 $^r$ & -1.05            &      -           & 0.143     & 1.04       \\
  M 3-8     & 1.93    & 0.860       & -            & 3.73        & 3.2      &              &           & -1.48            &     4.50         & 0.095     & 0.53       \\
\vspace{0.1cm}
  H 1-40    & 2.52    & 0.939       & -            & 4.26        & 3.0      & 20 $^{GZ,R}$ & 1000      & -1.23            &      -           & 0.044     & 0.66       \\
  M 2-23    & 1.28    & 0.924       & -            & 4.00        & 1.0      & 14 $^{R}$    & 1500 $^r$ & -0.15            &     4.73         & 0.108     & 0.02       \\
  He 2-260  & 0.76    & 0.012       & -            & 4.16        & 1.0      & 17 $^{G}$    & 3500      & -0.65            &     4.41         & {\em 0.2} & {\em 0.74} \\
  H 1-32    & 1.63    & 0.787       & -            & 3.74        & 1.0      & 11 $^{R}$    & 1900 $^r$ & -1.37            &      -           & 0.012     & 0.07       \\
  H 1-35    & 1.33    & 0.525       & 0.002        & 4.26        & 1.5      & 10 $^{GZ}$   & 3100      & -0.26            &     4.65         & 0.103     & 0.20       \\
\hline
\hline
\end{tabular}
\end{minipage}

{\em References}: $ RRA $ - \cite{Robinson1982},
                  $ GZ  $ - \cite{Gesicki2000},
                  $ G $ - \cite{Gesicki2006},
                  $ R $ - \cite{Richer2008}.\\
{\em Note}: Kinematic ages {\em{t}}$_{kin}$ are from \cite{Gesicki2000} and 
            \cite{Gesicki2006} or calculated assuming distance 8.5\,kpc ($r$) or 
            from Shklovski method ($s$).

\end{table*}

\section{Nebular properties and evolutionary status}

We analyze different nebular properties that can shed the light on the three
groups of bulge PNe and allow us to infer their evolutionary status and
origin. Since some of them (in particular \DCcr) have infrared spectra
resembling those of PNe with emission-line CSs, we also check whether any
other properties would associate them with such objects.

In Table~2 we report the basic nebular parameters of the investigated PNe.
Most of them were adopted from \cite{Gorny2009}, recomputed using the
methods and formulae from \cite{SS1999}, or adopted from other literature
sources.

\subsection{Bulge location}

\begin{figure}
\resizebox{\hsize}{!}{\includegraphics{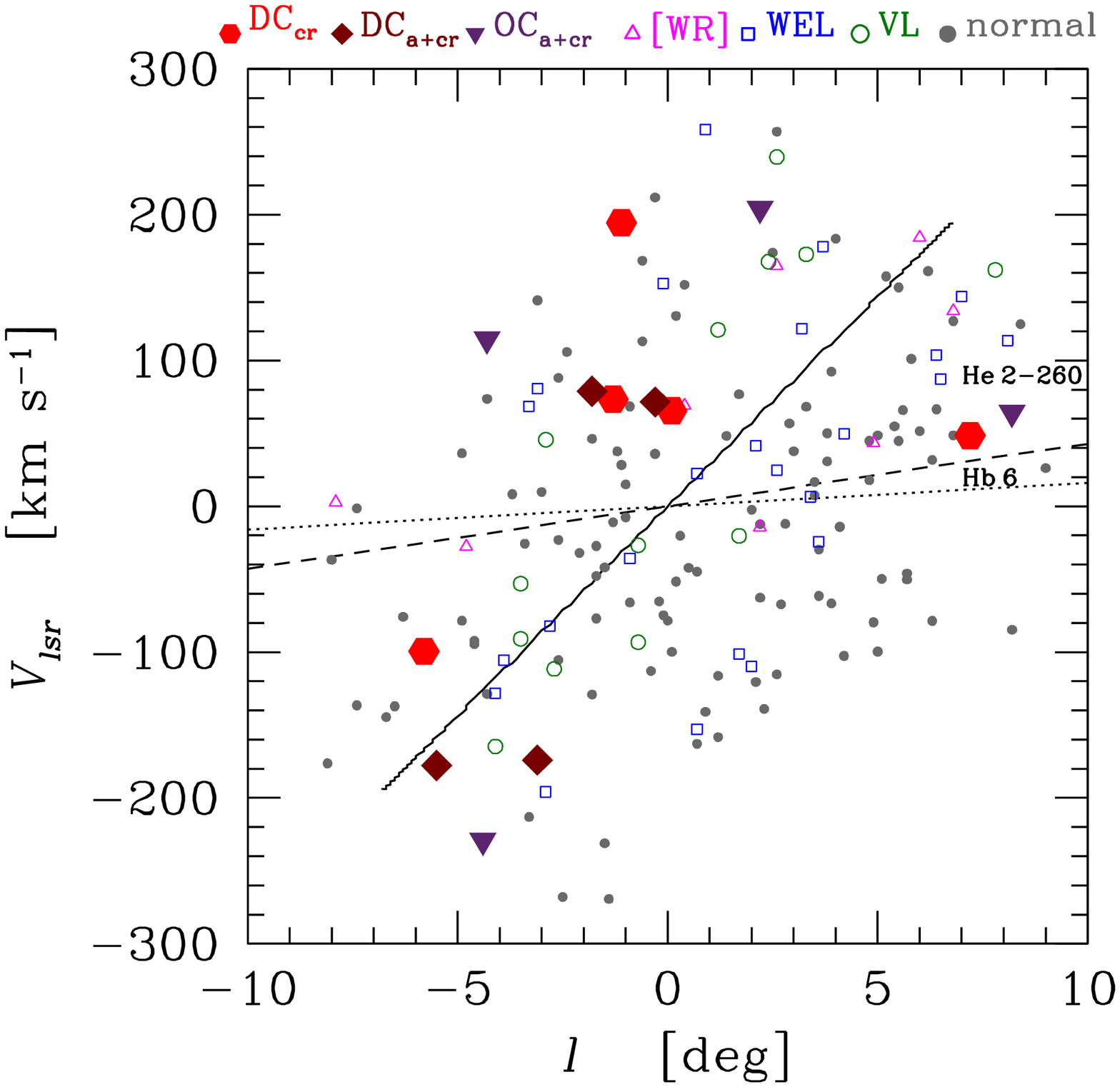}}
\caption[]{ 
  Radial velocity versus Galactic longitude coordinate of the investigated
  PNe.  The big filled symbols mark: \DCcr\ as red hexagons, \DCacr\ brown
  diamonds, and \OCacr\ as violet reversed triangles.  Galactic bulge PNe
  with emission-line CSs are marked with smaller open symbols: [WR] as pink
  triangles, WEL as blue squares, and VL as green circles.  {\em Normal}
  Galactic bulge PNe are presented with small dark grey dots.  Rotation
  curves for PNe at 1, 4, and 6\,kpc circular orbits assuming Galactocentric
  rotation velocities of 220\,km/s are plotted with solid, dashed, and
  dotted lines respectively.
}
\label{l_vlsr}
\end{figure}

 As mentioned in Sect.~2, most of the PNe analyzed here are not only simply
observed towards the center of the Milky Way but most probably physically
pertain to the Galactic bulge. In particular, they are located less than
10\degr\ from the center of the Galaxy. Figure~\ref{l_vlsr} presents the
radial velocities {\em{V}}$_{lsr}$ of bulge PNe corrected for solar
motion\footnote{
 The radial velocities used in this paper have been taken from 
 \cite{Durand1998} and corrected to the local standard of rest using the
 formulae of \cite{Beaulieu2000}} 
as a function of Galactic longitude coordinate {\em{l}}.
Although the samples of \DCcr, \DCacr, and \OCacr\ are not numerous, we 
can state that members of each of them are found at different longitudes 
within the bulge and no grouping at special locations can be 
distinguished\footnote{
 Recently \cite{Gorny2009} noticed a difference in locations between bulge
 [WR], VL, and WEL~PNe with the last two groups typically being located at
 longitudes less than 4{\fdg}5 from the center whereas bulge [WR]~PNe are
 often found at greater longitudes. Obviously, the 4{\fdg}5 longitude does
 not seem to be important for any of the subsamples investigated here.}.

 As can be noted in \figref{l_vlsr}, the velocities of Galactic bulge \DCcr,
\DCacr, and \OCacr\ objects (large filled symbols) are usually large or very
large and therefore their kinematic properties are typical of PNe
physically related to the bulge system (compare with Fig.\,13 of
\cite{Gorny2004}). On the other hand, the two objects already assigned to
Galactic disk (Hb\,6 and He\,2-260) have small {\em{V}}$_{lsr}$
velocities that are characteristic of that PNe population.

\begin{figure}
\resizebox{\hsize}{!}{\includegraphics{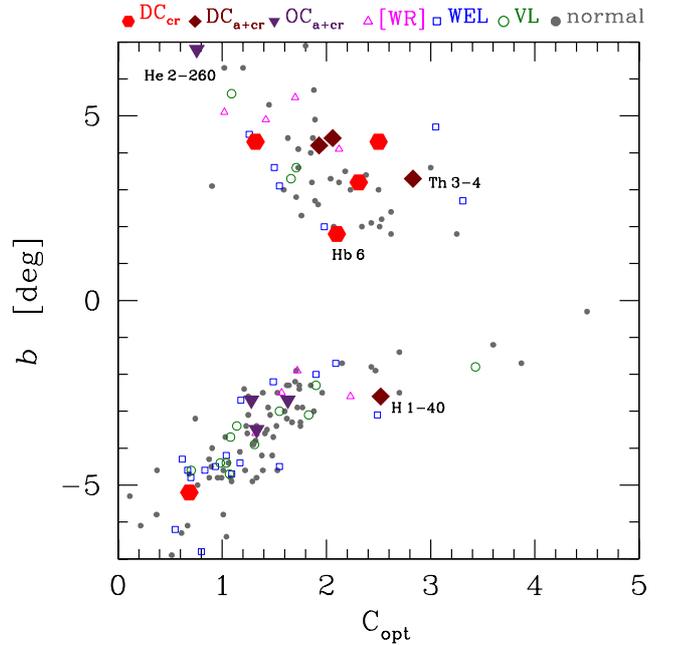}}
\caption[]{ 
  Locations of investigated PNe in the 
  Galactic latitude coordinate
  versus logarithmic extinction at H$\beta$ parameter C$_{opt}$. The meaning of
  the symbols is the same as in \figref{l_vlsr}. 
}
\label{C_b}
\end{figure}

For the other Galactic coordinate, the latitude {\em{b}}, the bulge PNe
analyzed here show nothing peculiar in their distribution. They are found at
{\em{b}} smaller than about 5\degr, but, as for all the other known PNe,
avoid latitudes below 2\degr. This is simply because of the interstellar
dust prohibiting their discovery. However, in \figref{C_b} we plot
the bulge PNe in {\em{b}} coordinates versus logarithmic extinction parameter
at H$\beta$ (C$_{opt}$) and some display a very interesting
property: the members of \DCacr\ (in particular Th\,3-4 and
H\,1-40) have much greater extinction than expected of typical PNe
located at their latitudes off the Galactic plane.

The extinction plotted in \figref{C_b} (see also Col.\,2 of Table~2) were
derived from the ratios of hydrogen lines in optical spectra. This was
accomplished mainly by comparing the observed Balmer H$\alpha$/H$\beta$
ratio with its theoretically expected value, although sometimes
H$\alpha$/H$\gamma$ had to be used \citep[see details in][]{Gorny2009}. We
checked that when extinction can be derived from both of these
ratios the agreement is usually very good.

The value of PNe extinction may be above average when the matter along
the line of sight has a different blocking properties or if the object is
simply located behind a larger amount of interstellar dust. This is the most
straightforward explanation, although the chances of finding all four
\DCacr\ with these conditions can be evaluated at only a few percent. 
Therefore at least partly the excessive extinction towards \DCacr\ could
also be attributed to some specific properties of their own dust or the dust
nearby around them. In that way, \DCacr\ would be interesting candidates for
studying the internal extinction of PNe.

\begin{figure}
\resizebox{\hsize}{!}{\includegraphics{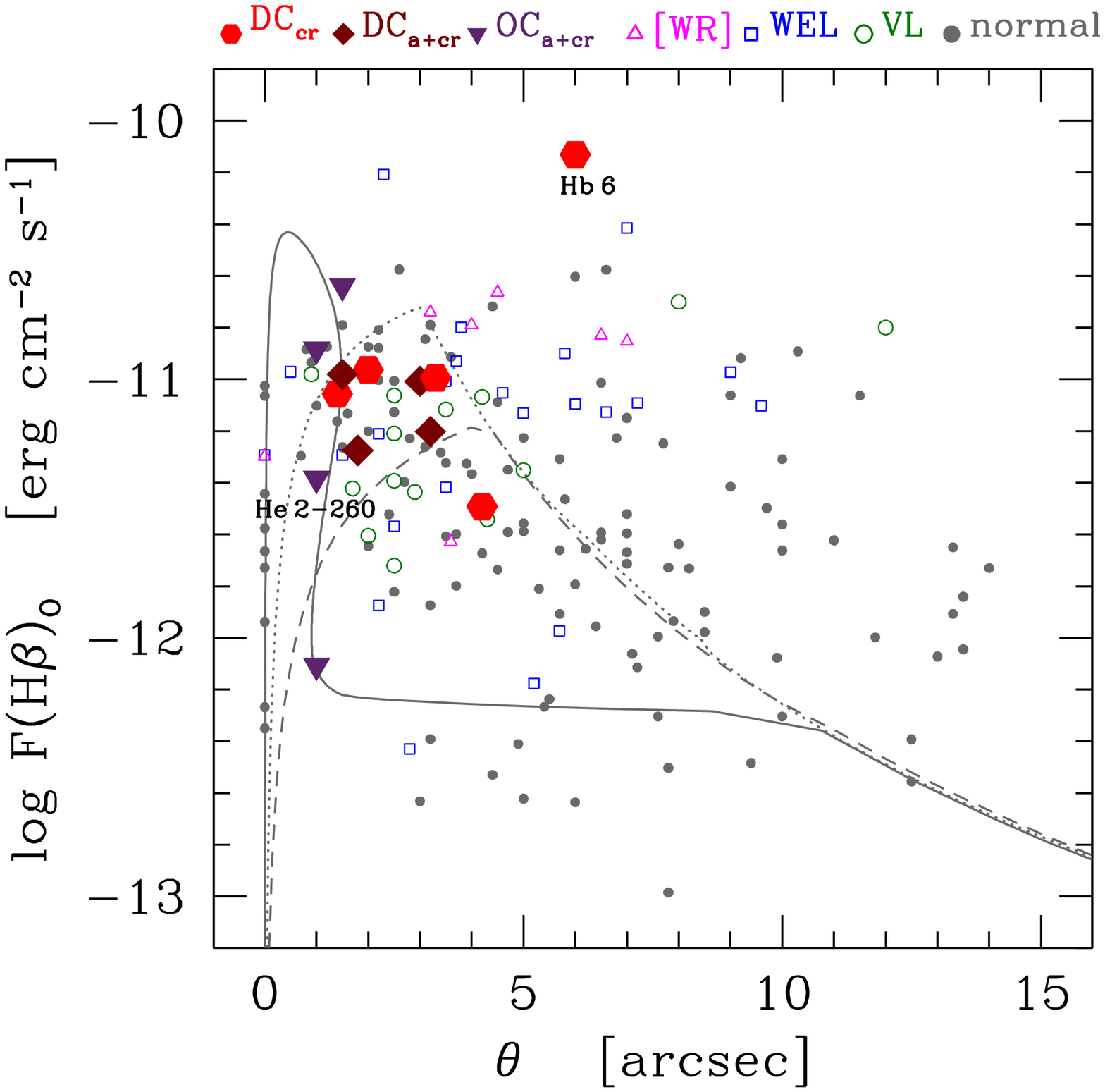}}
\caption[]{
  The relation between apparent diameter and reddening-corrected H$\beta$ flux
  for Galactic bulge PNe. The lines present model calculations
  for CSs of 0.57\Msun\ (dashed lines), 0.60\Msun\ (dotted), and
  0.64\Msun\ (solid), adopting the total nebular mass of 0.20\Msun.
  The meaning of the symbols is the same as in \figref{l_vlsr}.
}
\label{diam_fhb}
\end{figure}

\subsection{Diameters and densities}

In \figref{diam_fhb}, the reddening-corrected fluxes of the hydrogen
H$\beta$ line of GBPNe are presented as a function of apparent angular
diameters (Col.\,6 in Table~2). As a reference we overplotted the
theoretical tracks for PNe at a distance of 8.5\,kpc with CSs of
three different masses 0.57, 0.60, and 0.64\Msun, evolving according to the
predictions of \cite{Bloecker1995} and assuming they radiate like a
black body. The surrounding nebula were described by a simple model of
uniformly filled sphere with total gas mass of 0.2\Msun, filling factor
$\epsilon$=0.75, and expansion at the constant velocity of 20\,km/s.  As can
be seen in this figure, the
\DCcr\ and \DCacr\ PNe occupy a rather restricted region of the plot. Their
H$\beta$ fluxes are typically 2 times lower than those of the majority of
[WR] PNe in the bulge, comparable to the brightest of WEL PNe and usually
brighter than VL PNe. The diameters of \DCcr\ and \DCacr\ PNe range
from about 1\arcsec to 5\arcsec, suggesting that they are young or the
nebulae are expanding slowly. However, expansion velocities are known for
most of these objects (see Col.\,7 of Table\,2) and seem normal with a
typical value of 20\,km/s. Using the data collected in \cite{Gesicki2007},
one can check that for the remaining GBPNe, \normal, [WR], WEL
and VL PNe combined (37 entries in their Table\,3), the expansion velocities
have a rather flat distribution from 10 to 32\,km/s with a median value of
V$_{exp}$=22\,km/s.

The diameters of \OCacr\ PNe are typically only 1\arcsec, which are the
smallest among known GBPNe. However there are indications, that their
expansion velocities may be smaller than average as they range from 10\,km/s
to only 17\,km/s.

Assuming a simple model of constant nebular expansion that starts when the
object leaves the AGB and knowing the angular diameter, distance, and
V$_{exp}$, it is possible to derive the kinematic ages {\em{t}}$_{kin}$ of
PNe.  They are listed in Col.\,7 of Table\,2 and range from 500 to 3800
years. For comparison, the median age of GBPNe is 2900 years using data
collected in \cite{Gesicki2007}\footnote{
 This median age of GBPNe may be underestimated.
 There is a possibility of important selection effects in the sample of
 \cite{Gesicki2007} that seems
 strongly biased towards PNe with small angular diameters as the median
 $\theta$ is 3$\farcs$5 in their sample compared with 5$\farcs$4 of our \normal\ PNe.
}. This means that the objects analyzed here may not be extremely young but
clearly belong to the younger PNe in the Galactic bulge. This is
unsurprising taking into account the possible selection effects when choosing
objects for Spitzer observations, as described in Sect.\,2.

\begin{figure}
\resizebox{\hsize}{!}{\includegraphics{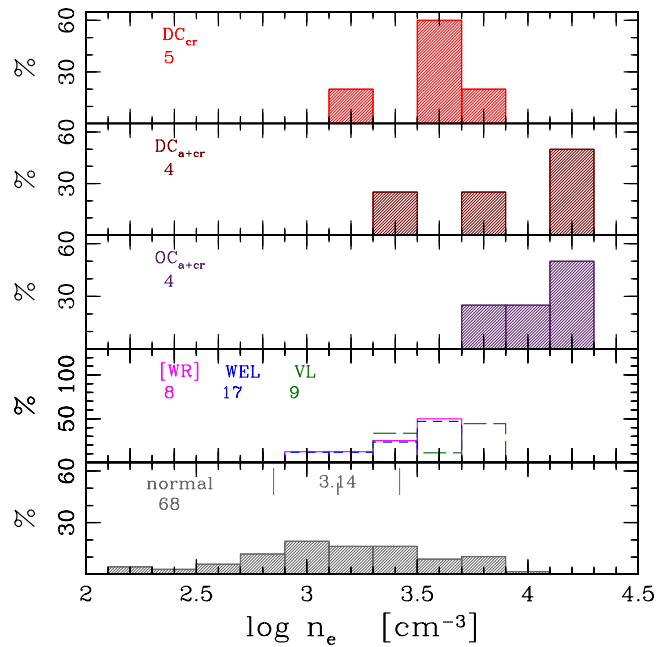}}
\caption[]{
  Distributions of electron densities for the different groups of Galactic bulge
  PNe.  For the \normal\ PNe (bottom) the median value along with the 25 and
  75 percentiles are marked with three short vertical lines above the
  histogram. Total numbers of objects included are shown in the left-hand
  parts of the panels below sample names.
}
\label{hist_Ne}
\end{figure}

In \figref{hist_Ne}, we present the distributions of electron densities
derived from [SII] 6717/31\AA\ line ratio for the different groups of GBPNe.
The values of n$_e$ for individual objects can be found in Col.\,5 of
Table~2\footnote{Table~4 is available online.}. Electron density is another
independent parameter related to the age of the object. Surprisingly, as can
be seen in \figref{hist_Ne}, objects analyzed in this paper are among the
densest of PNe in the Galactic bulge region. In particular, \OCacr\ and
\DCacr\ PNe, i.e., the objects with the 10\um\ emission feature are much
denser than the median log\,n$_e$=3.14 of \normal\ GBPNe. The fact that
they are also denser than the 75 percentile value of \normal\ PNe
implies that the distributions are truly different. The \OCacr\ and
\DCacr\ are also considerably denser than the PNe with emission-line CSs.

\subsection{Evolutionary status}

The evolutionary state of the central star of the planetary nebula is best
described by its temperature. In Col.\,10 of Table~2, we present the stellar
temperatures estimated with the Zanstra hydrogen method.  Unfortunately, the
data necessary to derive temperatures are not available for many \DCcr\ and
\DCacr\ objects and it is then difficult to compare them with the
temperatures of the other GBPNe.

However, the most characteristic features of the planetary nebula spectrum
are emission lines of many different atoms and their ions at different
levels of excitation. As the temperature of the CS increases, the
lines of increasingly higher ionization stages become observable. We
take advantage of this property to at least qualitatively investigate the
evolutionary stage of all the objects, including those with unknown
CS temperature. For this purpose we divide them into three
excitation classes: i) those characterized by nebulae in which most of the
oxygen atoms remain in the form of O$^{+}$ ions i.e.,
O$^{++}$/(O$^{+}$+O$^{++}$)$<$0.30 (for the coolest CSs); ii) those with
most of their nebular oxygen in the form of O$^{++}$ but no or very little
He$^{++}$ ions present i.e., O$^{++}$/(O$^{+}$+O$^{++}$)$>$0.30 and
He$^{++}$/(He$^{+}$+He$^{++}$)$<$0.03 (for the intermediate temperature
CSs); iii) the remaining objects with considerable amounts of helium
He$^{++}$ ions i.e., He$^{++}$/(He$^{+}$+He$^{++}$)$>$0.03 (for the hottest
CSs). Using the three excitation classes so defined, we checked separately
for each group of PNe discussed in this work the distributions of H$\beta$
surface brightness. The parameters O$^{++}$/(O$^{+}$+O$^{++}$) and
He$^{++}$/(He$^{+}$+He$^{++}$) for individual \DCcr, \DCacr, and \OCacr\ PNe
are given in Cols.\,4 and 5 of Table~2 and the values of S$_{H\beta}$ in
Col.\,9 of that Table.

\begin{figure*}
\resizebox{0.32\hsize}{!}{\includegraphics{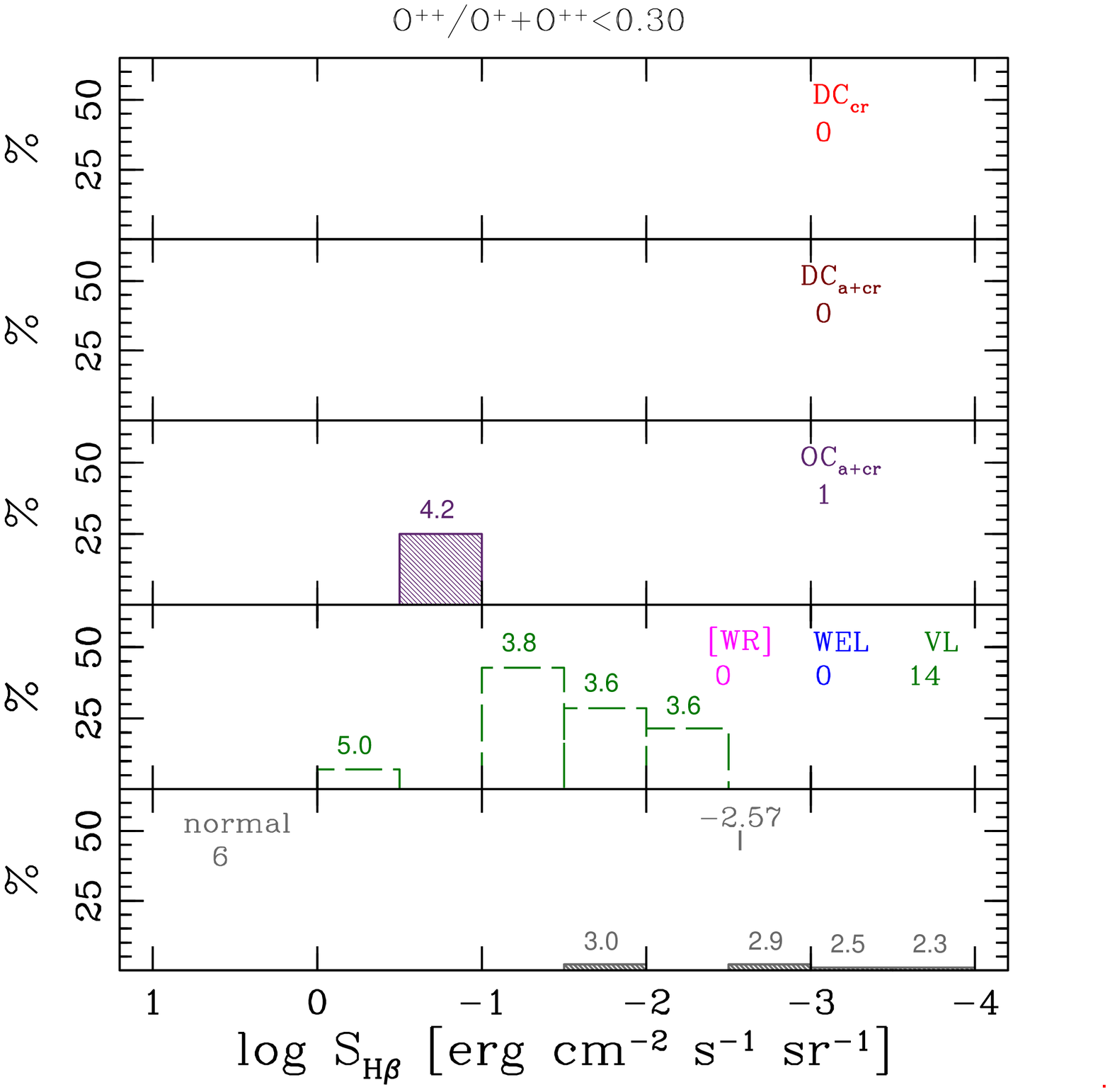}}
\resizebox{0.32\hsize}{!}{\includegraphics{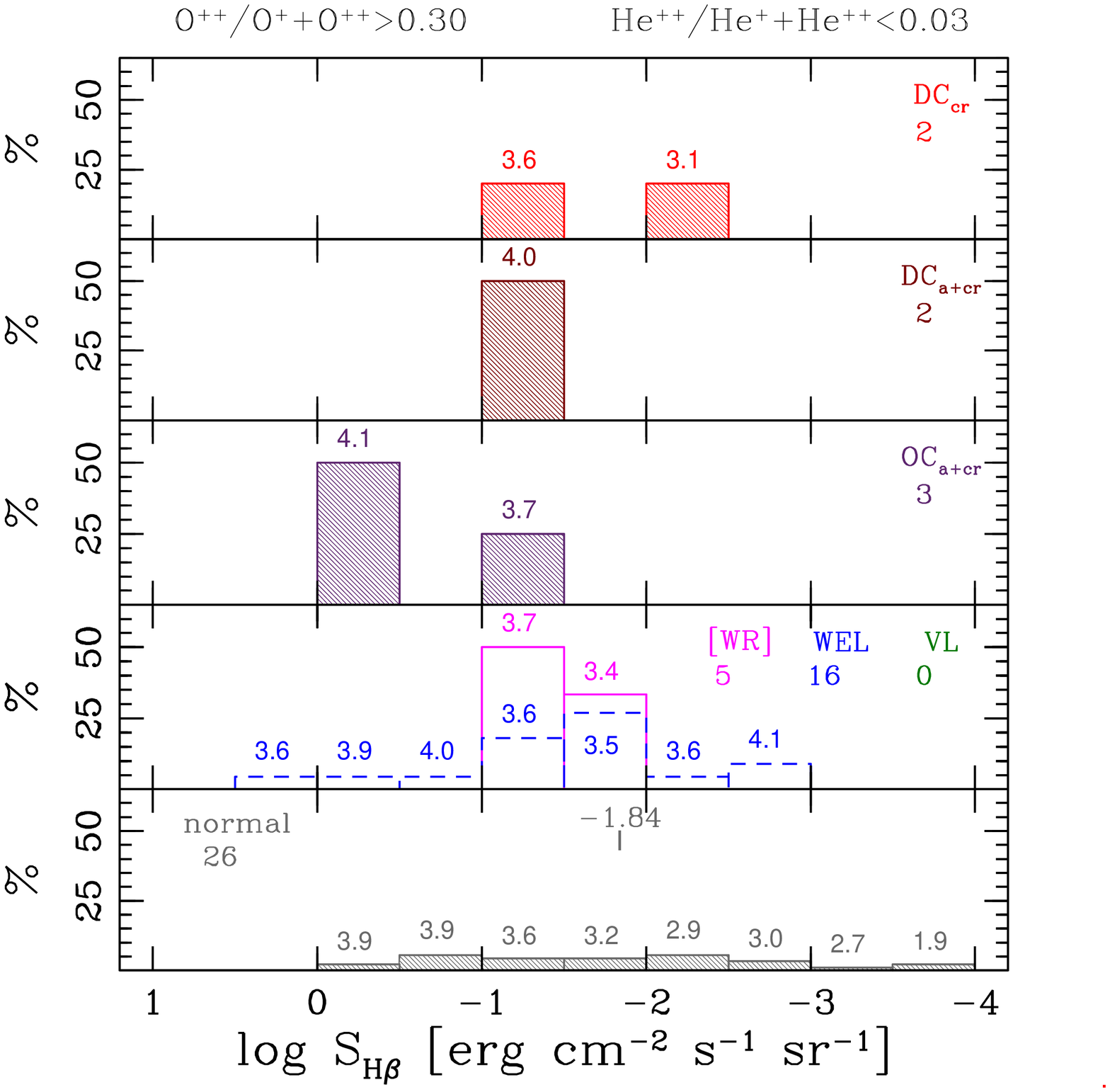}}
\resizebox{0.32\hsize}{!}{\includegraphics{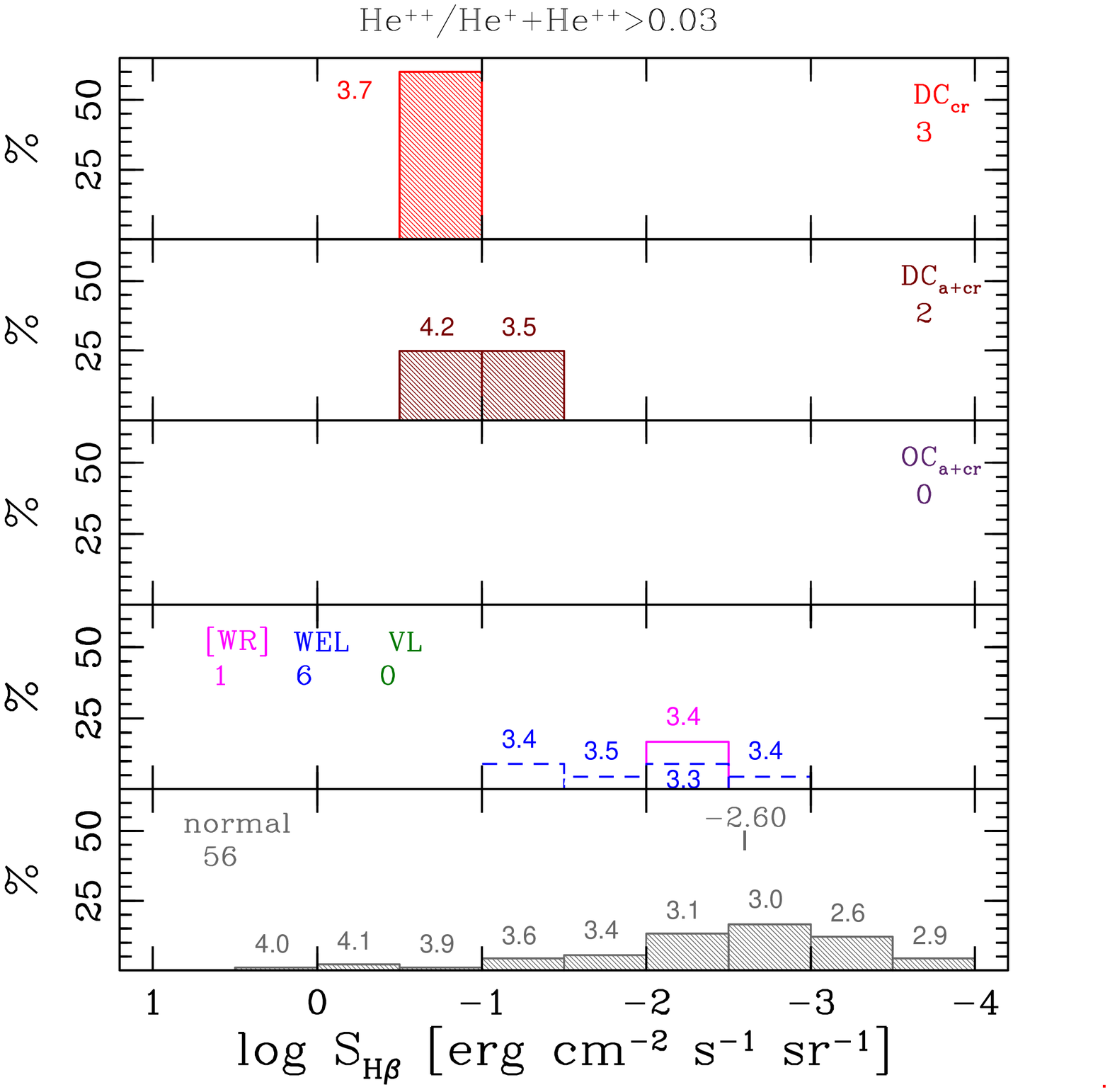}}
\caption[]{
  The distribution of S$_{H\beta}$ surface brightness for different types of
  GBPNe. Separate histograms have been constructed for objects with
  different ionization classes as described at the top of the panels. Above
  each bar the logarithm of the mean electron density of the PNe comprising
  it is given (in cm$^{-3}$).
}
\label{hist_shb}
\end{figure*}

The S$_{H\beta}$ parameter constitutes a good measure of the evolutionary
advancement of the nebula and changes by a few orders of magnitude between
the formation of the observable nebula and the moment it disperses into the
interstellar medium. In \figref{hist_shb}, we present the histograms of
S$_{H\beta}$ for PNe with the coolest, intermediate, and hottest CSs in the
left, middle, and right panels respectively. Above each histogram, the average
nebular electron densities of the PNe that comprise the bars are
overplotted.

\begin{figure*}
\resizebox{0.32\hsize}{!}{\includegraphics{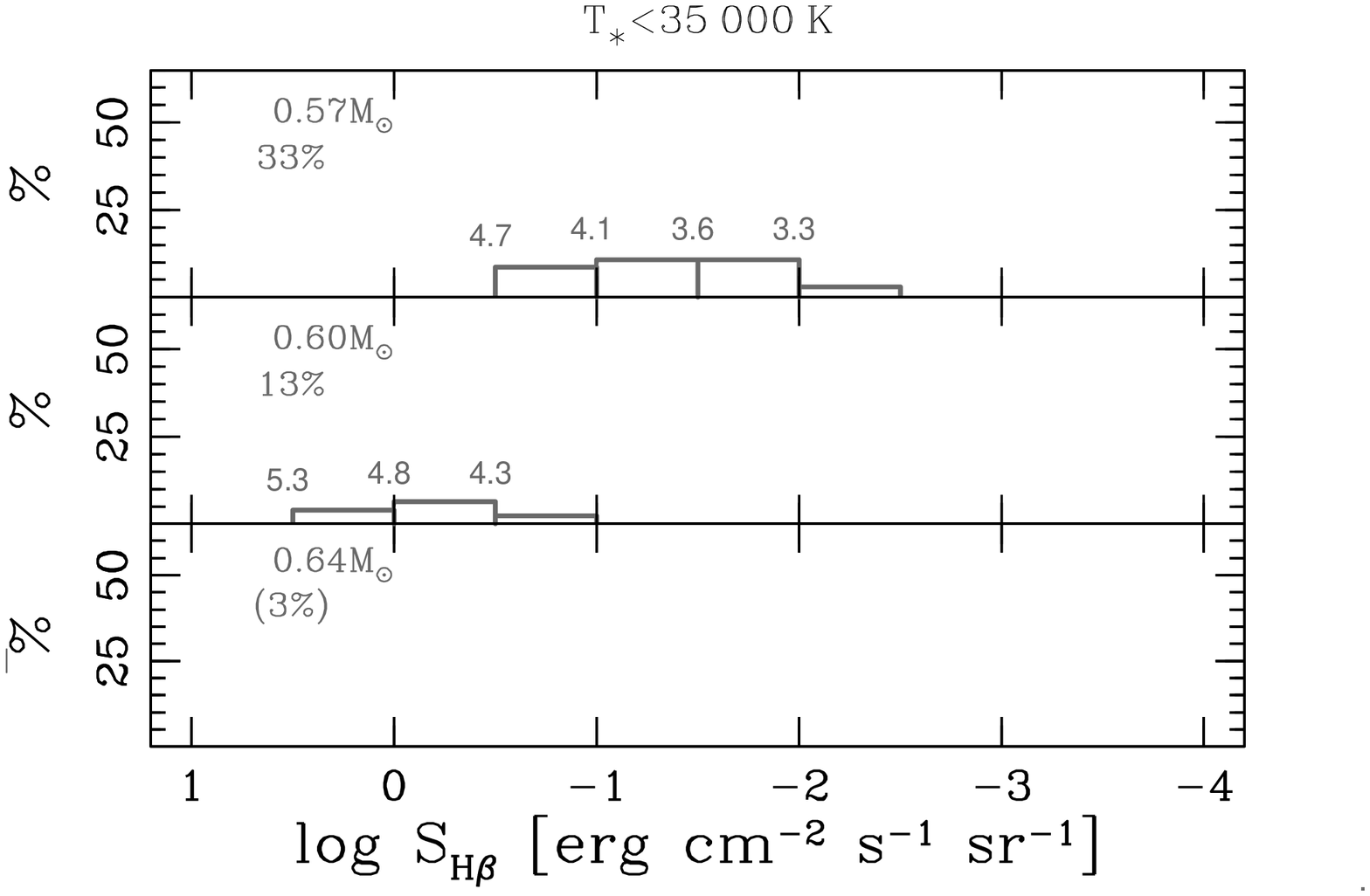}}
\resizebox{0.32\hsize}{!}{\includegraphics{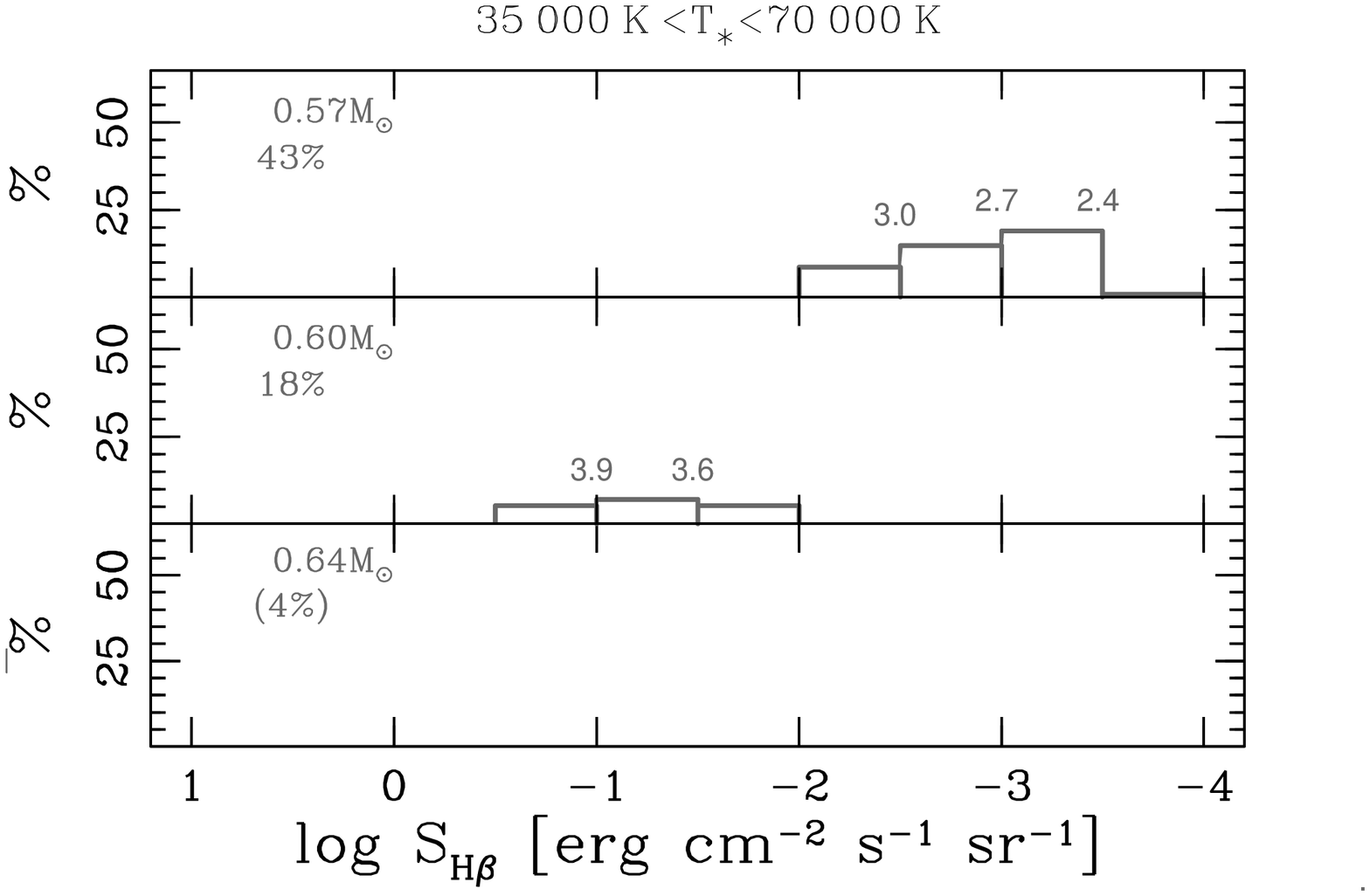}}
\resizebox{0.32\hsize}{!}{\includegraphics{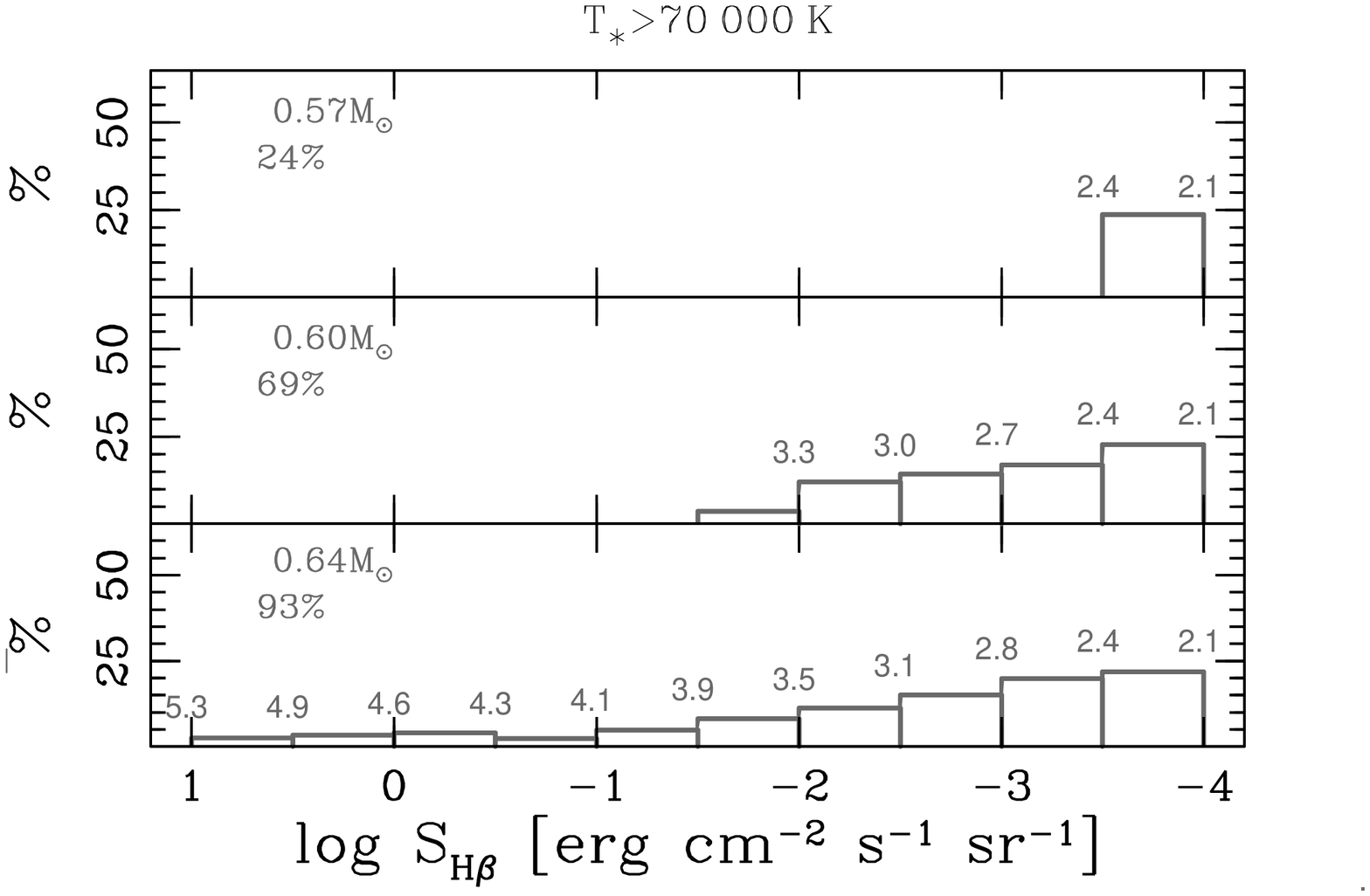}}
\caption[]{
  Theoretical prediction of S$_{H\beta}$ distributions for PNe with total
  gas mass of 0.2\Msun,
  parameter $\epsilon$=0.75, 
  and expansion at 20\,km/s for the CSs of 0.57, 
  0.60, and 0.64\Msun. Separate histograms are presented for objects with 
  CS temperature T$_{\star}<$35kK (left panel), 
  35kK$<$T$_{\star}<$70kK (middle), and T$_{\star}>$70kK (right). 
  The nebular electron densities are in addition plotted above the bar
  limits as expected from model calculations for PNe characterized by a given 
  S$_{H\beta}$ value (log n$_e$ in cm$^{-3}$). 
  The numbers under CS mass give the cumulative
  percentage of objects expected in the analyzed temperature domain (in
  parenthesis if S$_{H\beta}$ is expected larger than the presented range of
  values).
}
\label{hist_shb2}
\end{figure*}

In \figref{hist_shb2}, we present for comparison the theoretical predictions
for the histograms of S$_{H\beta}$. We used
the same models as already presented in Sect.\,4.2 and
\figref{diam_fhb}. We considered PNe with CSs of three
different masses, 0.57, 0.60, and 0.64\Msun\ evolving according to the
calculations of
\cite{Bloecker1995}. The same simple model of surrounding nebula were
assumed of uniformly filled sphere with total gas mass of 0.2\Msun, filling
factor $\epsilon$=0.75, and expanding at the constant velocity of 20\,km/s.
To qualitatively reconstruct the ionization conditions applied above to the
real observed PNe, we divided the life of the model objects into three
periods with CS temperature below 35\,kK, between 35\,kK and
70\,kK,
and finally above 70\,kK (left, middle, and right panels in
\figref{hist_shb2}, respectively). The height of each bar is directly
proportional to the period the object is characterized by the given
S$_{H\beta}$. It is normalized by the total time nebula should be bright
enough to be detected (log\,F(H$\beta$)$>$-12.6) but not too extended and
dispersed ($\Theta$$<$14arcsec)\footnote{
 Compare with the actual values for bulge PNe in \figref{diam_fhb}
 }.  
The nebular electron densities expected from model calculations are also
overplotted for the S$_{H\beta}$ starting and ending limit of each histogram
bar.
 
Analyzing the distributions of \DCcr\ and \DCacr\ in \figref{hist_shb}, we
note that PNe of both groups seem to be at a similar stage of nebular
evolution. Roughly half of their population seems to be associated with CSs
of intermediate temperatures, while the other half has CSs that are already
very hot. There are no \DCcr\ nor \DCacr\ PNe with cool CSs but the apparent
deficit of objects with cool CSs seems to be a general property
of GBPNe (with a clear exception of the VL PNe).

The \DCcr\ and \DCacr\ with stellar temperatures in the range 35kK to 70kK
are well reproduced by models with intermediate mass CSs (compare with
predictions for 0.60\Msun\ star in \figref{hist_shb2}). It is also important
to note that their derived S$_{H\beta}$ have values typical of other
GBPNe, including objects with emission-line CSs. Therefore, these \DCcr\ and
\DCacr\ PNe do not seem to be less evolutionary advanced than [WR] or WEL
PNe and thus cannot be their predecessors\footnote{
 The existence of such predecessors
 with undiscovered or not yet active strong stellar winds could be
 one of the explanations why Galactic bulge [WR] PNe have only intermediate
 [WC] class CSs.}.

A different situation can be seen in the right panel of \figref{hist_shb},
since the S$_{H\beta}$ of \DCcr\ and \DCacr\ are clearly larger than those
of both \normal\ PNe and of the few GBPNe with hot emission-line CSs.
By comparing with the models in \figref{hist_shb2}, it is apparent that \DCcr\ and
\DCacr\ PNe with CS temperatures above 70\,kK originate in more
massive AGB progenitors than the group discussed above. Analyzing the model
distributions, it can also be deduced that they should have a considerable
number of unrecognized ancestors among \normal\ PNe with hot CSs. The
evolutionary link to some PNe with emission-line CSs also cannot be
excluded\footnote{ 
 The analysis is hampered by the 
 \cite{Bloecker1995} tracks being applicable directly only to PNe with
 H-burning nuclei, i.e., not to [WR] type CSs.}.

In summary, it seems that \DCcr\ and \DCacr\ PNe can be found among PNe with
both massive and intermediate-mass CSs. As both groups are characterized by
similar range of temperatures and S$_{H\beta}$ values but have clearly
different electron densities (see \figref{hist_Ne}), their evolutionary
status cannot be identical. The details may be derived by future
complete modeling. However, the higher electron densities of
\DCacr\ PNe could be explained if the distribution of gas was more clumpy in
these PNe.

\begin{figure*}
\resizebox{0.32\hsize}{!}{\includegraphics{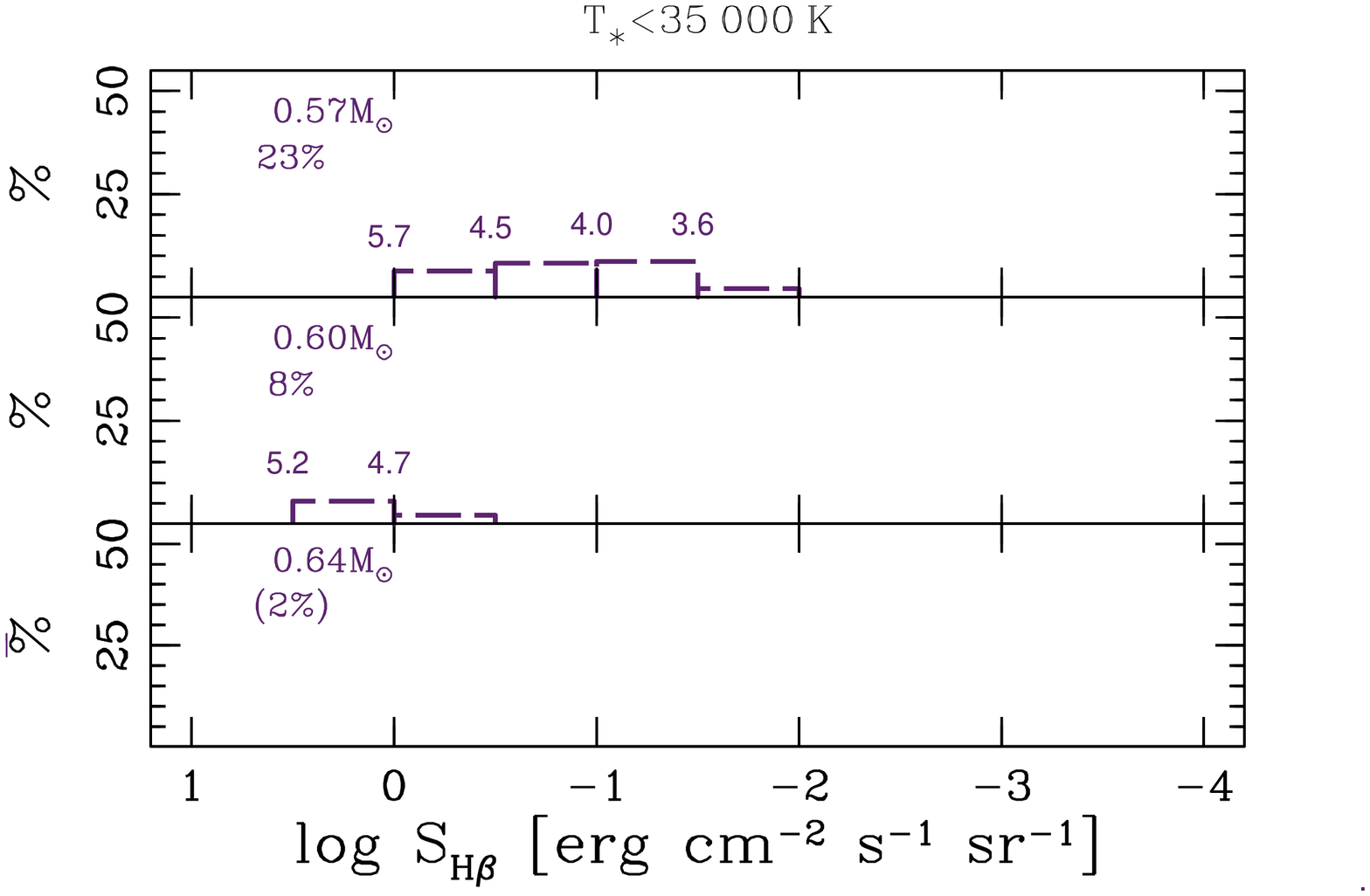}}
\resizebox{0.32\hsize}{!}{\includegraphics{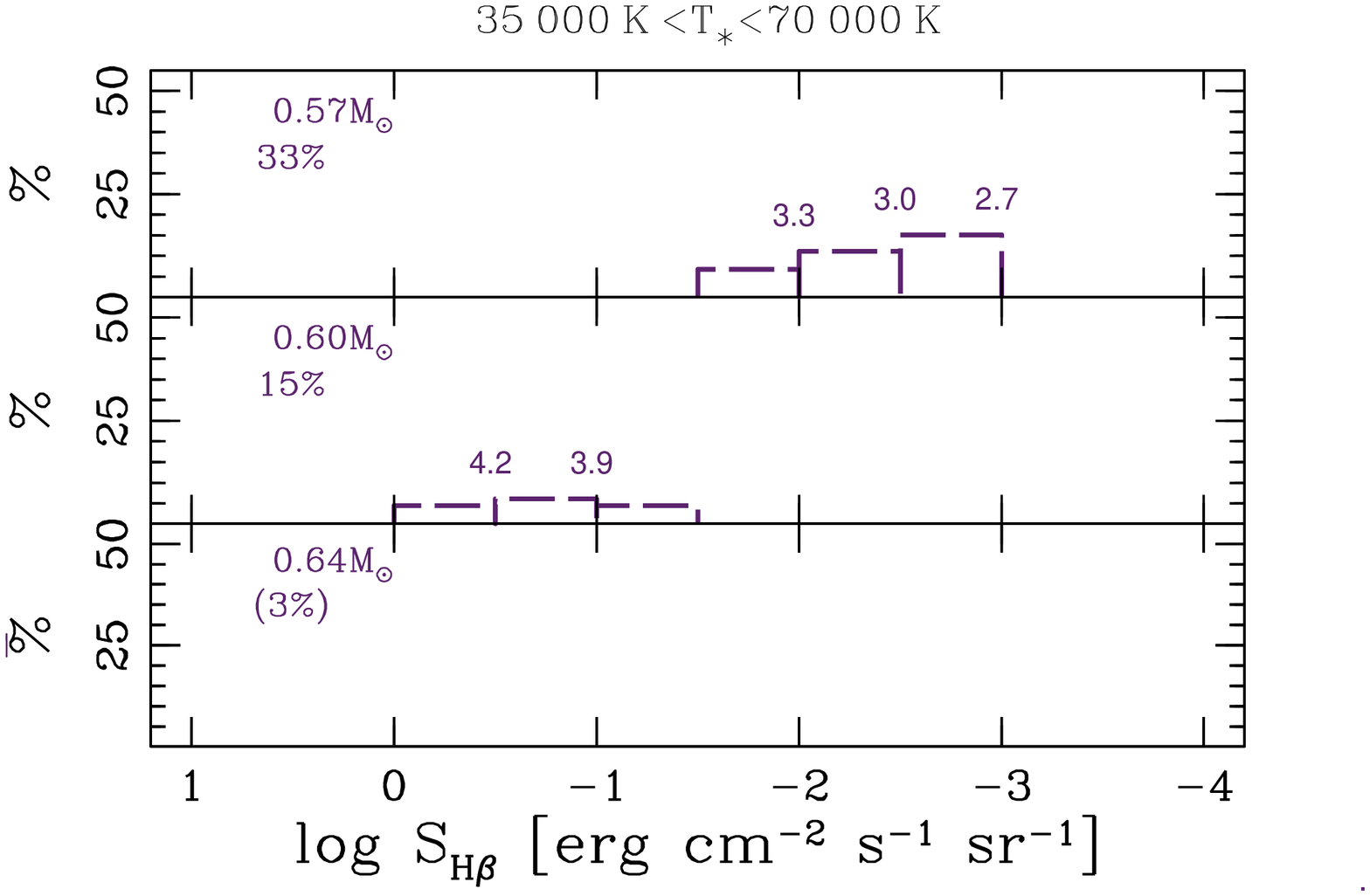}}
\resizebox{0.32\hsize}{!}{\includegraphics{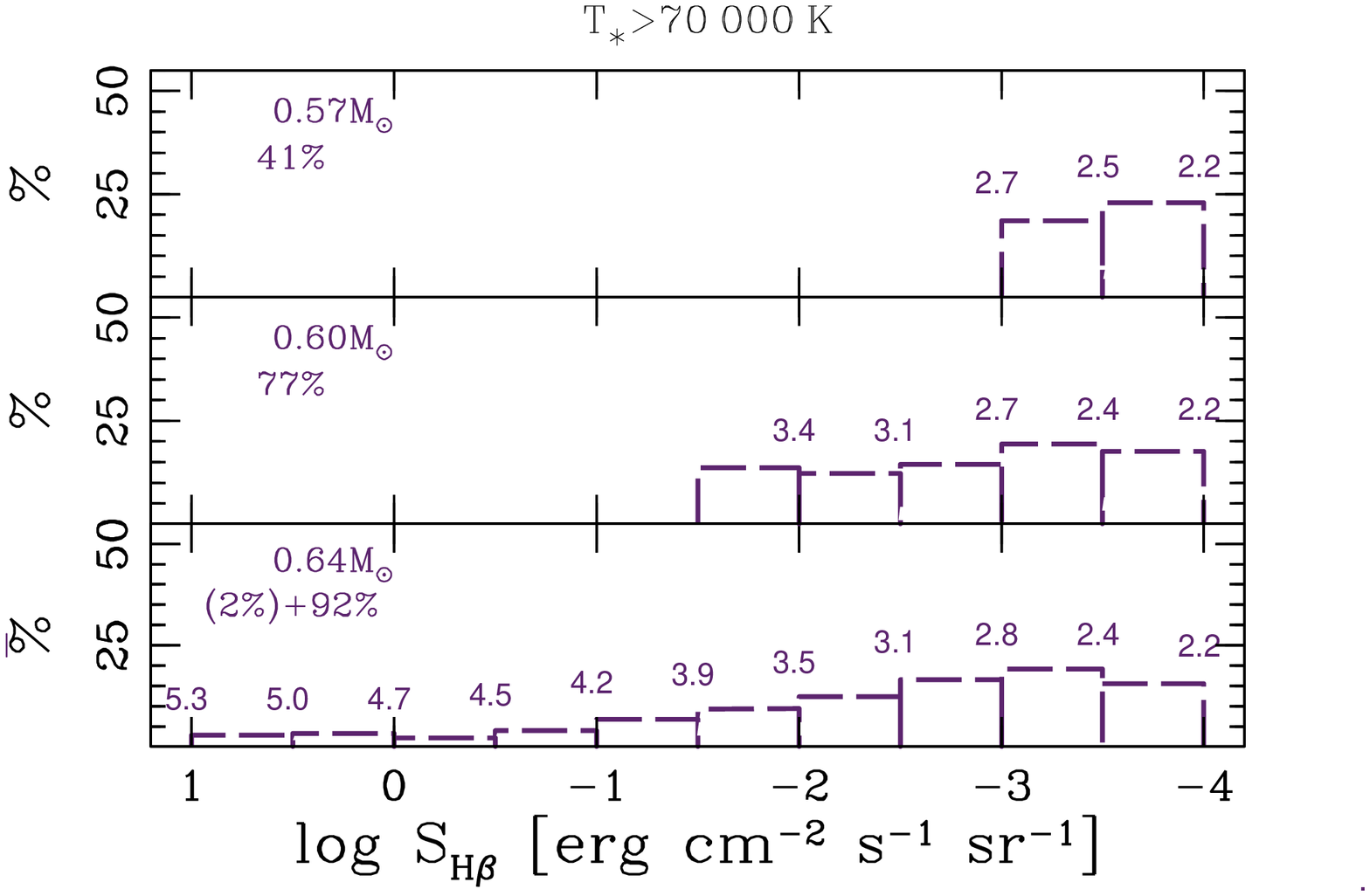}}
\caption[]{
  Theoretical prediction of S$_{H\beta}$ distributions for PNe of 0.1\Msun\
  and expanding at 12\,km/s in the case of central stars of 0.57, 0.60, and
  0.64\Msun. The separate histograms are presented for objects with central
  star temperatures T$_{\star}<$35kK (left panel), 35kK$<$T$_{\star}<$70kK
  (middle) and T$_{\star}>$70kK (right).  
}
\label{hist_shb3}
\end{figure*}

The S$_{H\beta}$ distributions of \OCacr\ PNe can be compared with the
separate theoretical predictions in
\figref{hist_shb3}\footnote{\figref{hist_shb3} is available online.}. Since
the \OCacr\ PNe evolve more slowly, the expansion velocity of 12km/s was
assumed in these models (see in Sect.\,4.4 and Table~2). In addition, a
total nebular mass of 0.10\Msun\ was adopted so the models fit the small
diameters of these PNe and do not exceed the measured electron densities. A
qualitative agreement can be observed for \OCacr\ PNe with models of
intermediate-mass CSs while models assuming lower-mass progenitors can be
rejected. The only exception is He\,2-260, an object with the coolest CS and
a nebula with the fastest expansion among
\OCacr\ PNe that is probably located in the Galactic disk.

\subsection{Masses of nebular gas and dust} 

\begin{figure}
\resizebox{\hsize}{!}{\includegraphics{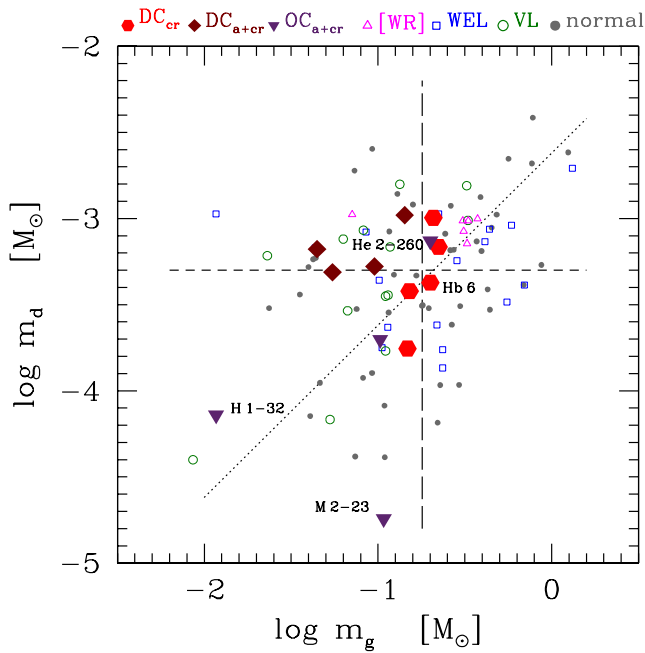}}
\caption[]{ 
  The mass of the dust versus the mass of the nebular gas for Galactic bulge
  PNe. The meaning of the symbols is the same as in \figref{l_vlsr}. The
  horizontal short-dashed and vertical long-dashed lines indicates,
  respectively, median {\em{m}}$_d$ and {\em{m}}$_g$ for \normal\ bulge PNe
  (small dark grey symbols) and the dotted line indicate typical
  {\em{m}}$_d$/{\em{m}}$_g$ ratio of these PNe.
}
\label{dg}
\end{figure}

Since the objects that we analyze were selected based on the appearance of
their infrared spectra, it is important to learn about their dust content.
The first important property is the dust-to-gas mass ratio
{\em{m}}$_d$/{\em{m}}$_g$. It can be easily estimated from observables by
adopting a simple model of the dusty nebula. We apply here the method from
\citet{SS1999} and recompute the parameters by adopting the new data on
electron densities derived by \cite{Gorny2009}. In addition, dealing with
Galactic bulge objects we can adopt the distance of 8.5\,kpc and derive
absolute values of nebular ionized gas and dust masses. Our results are
presented in \figref{dg} and in Cols.\,11 and 12 of Table~2. For the
Galactic disk PNe Hb\,6 and He\,2-260, the dust mass was calculated
from the derived {\em{m}}$_d$/{\em{m}}$_g$ ratio by assuming
{\em{m}}$_g$=0.2\Msun.

As can be seen in \figref{dg}, the dust content of \DCcr\ and \DCacr\ PNe is
usually similar to the typical value for \normal\ GBPNe (median
log\,{\em{m}}$_d$=-3.30, represented by horizontal short-dashed line in the
plot).  It can be recalled here that \DCacr\ can be suspected to have a
considerable internal extinction (Sect.\,4.1). However, since the dust
content in these PNe is not exceptionally greater than in other objects, it
could not explain excessive internal extinction. The reason for it would
have to be in some specific properties of their dust that allow it to block
more radiation than in other PNe.

As far as the dust-to-gas ratio of the PNe presented in \figref{dg} is
concerned, the \DCacr\ represent the group with the highest ratio (see
their location above the dotted line representing the median
{\em{m}}$_d$/{\em{m}}$_g$=2.47$\times$10$^{-3}$ relation of \normal\ PNe). The
\DCcr\ are much closer to this line meaning their {\em{m}}$_d$/{\em{m}}$_g$
ratios have normal values. However, we recall that if some part of the
nebula is still not ionized the value of {\em{m}}$_d$/{\em{m}}$_g$ that we
derive may be overestimated. This is e.g., most probably true for the VL
PNe that are in an extremely low ionization state \citep{Gorny2009}. For
\DCacr\ PNe, the derived ionized gas masses are all below
the median mass of \normal\ PNe in \figref{dg}
({\em{m}}$_g$=0.18\,\Msun, marked with vertical long-dashed
line). This indicates that the \DCacr\ PNe may also be only partially
ionized, if their total nebular masses are similar to those of
\normal\ PNe.

As can be seen in \figref{dg}, the mass of dust in individual \OCacr\ PNe
may differ by more than an order of magnitude. For two of them M\,2-23 and
H\,1-32, we derived very small values of {\em{m}}$_d$ below 10$^{-4}$\Msun.
On the other hand, except for M\,2-23, the \OCacr\ objects have
{\em{m}}$_d$/{\em{m}}$_g$ ratio close to the value typical of \normal\ PNe.
If these nebulae were only partially ionized the true
{\em{m}}$_d$/{\em{m}}$_g$ would be smaller than our estimate. Indeed, for
H\,1-32 the ionized gas mass is exceptionally low {\em{m}}$_g$=0.01\Msun.
For the other two bulge objects from the \OCacr\ group, the derived
{\em{m}}$_g$ is also only about one half of the typical mass of \normal\ PNe.

\subsection{Dust temperature and infrared excess}

\begin{figure}
\resizebox{\hsize}{!}{\includegraphics{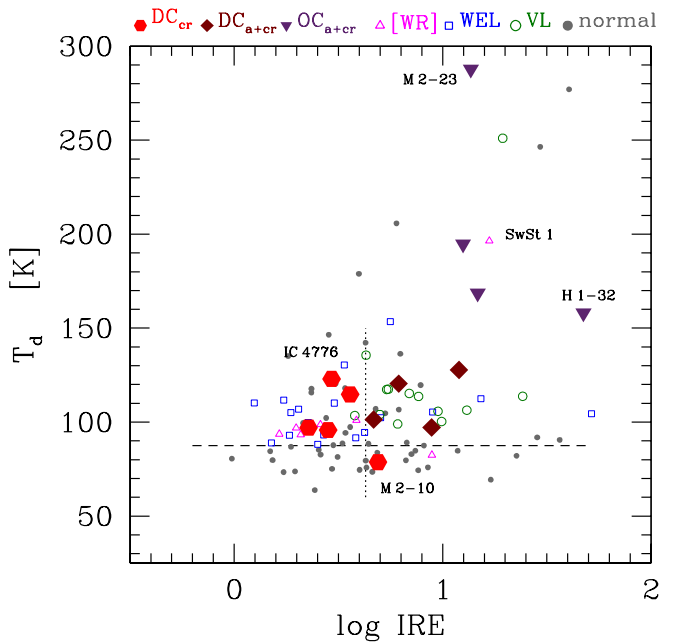}}
\caption[]{
  The dust temperature versus the infrared excess parameter for Galactic
  bulge PNe.  The meaning of the symbols is the same as in \figref{l_vlsr}.
  Dashed and dotted lines indicate median values for \normal\ bulge PNe
  (small black symbols).
}
\label{ire_Td}
\end{figure}

In \figref{ire_Td}, we present the dust color temperature T$_d$ derived from
the 25 and 60{\um} IRAS flux ratios versus infrared excess IRE defined as
\[
IRE = F_{IR} ~~/~~ 22.3 F_{H{\beta}}~,
\]
where total infrared flux F$_{IR}$ has been estimated by computing the
blackbody flux corresponding to the ratios of 25 to 60{\um}
fluxes of IRAS. The values of both T$_d$ and IRE were adopted from
\cite{SS1999}. Figure \ref{ire_Td} is found to be a good means of
distinguishing all the different groups of GBPNe. There are two lines in
this figure. The first represents the median IRE value for the
\normal\ GBPNe (dotted line) and the second the median T$_d$ for \normal\
PNe (dashed line). We note that almost all the nebulae that we analyze here,
\DCcr, \DCacr, and \OCacr, as well as PNe with emission-line CSs ([WR], WEL,
and VL) are located above the latter line, meaning they all have dust hotter
than the typical T$_d$ of \normal\ PNe. The possible reason for this is that
they are presumably at a relatively early or intermediate stage of PNe
evolution when the dust is hotter \citep{SS1999}.  In \figref{hist_shb} we
can see that the majority of evolved objects (with log\,S$_{H\beta}$$<$-2) are
the \normal\ PNe. In addition, selection effects may be important because
strong dust emission (i.e., selection criterion for Spitzer/IRS targets)
is more likely to be met in younger PNe with hotter dust when it is still
relatively close to the CS.

In \figref{ire_Td}, it is quite convincing that the \OCacr\ belong to the
objects with the hottest dust among bulge PNe.  Apart from the
above-mentioned relation with younger evolutionary age, other factors may be
playing a role. The T$_d$ can be higher in the case of low a
{\em{m}}$_d$/{\em{m}}$_g$ ratio in the nebula and when amorphous
silicate grains dominate the dust composition \citep[see Fig.\,7d,e
of][]{SS1999}. While there is only one \OCacr\ object with definitely very
low {\em{m}}$_d$/{\em{m}}$_g$ (M\,2-23, see in Sect.\,4.4) the members of
\OCacr\ are the only group of PNe analyzed here that show no clear signs of
the C-based dust and only evidence of the silicate dust grains are present
in their spectra.

In their models, \cite{SS1999} used the circumstellar silicates dominated by
amorphous grains, whereas in spectra of \DCcr\ PNe and GBPNe with
emission-line CSs there are clear signs of silicates in crystalline form
only. It is not certain how different forms of silicate grains may influence
the dust temperature. In \figref{ire_Td}, one can note that both IC\,4776
- an unusual, unique WEL PN with 10\um\ amorphous feature and SwSt\,1 - a
[WR] type PN with a similar property belong to the hottest of nebulae of
their respective types \citep[compare also with Fig.\,3 of][]{Gorny2001}. On
the other hand, the \DCacr\ PNe that also exhibit amorphous silicates
apparently have dust that is not convincingly hotter than for
\DCcr\ PNe that do not show the 10\um\ feature.

The vertical dotted line in \figref{ire_Td} indicates the median value of
log IRE for \normal\ PNe, which equals 0.63. This parameter and this
particular value are important because it appears to separate the different
groups of GBPNe. First, to the right of this line are \DCacr\ PNe (\OCacr\
are discussed below) and the peculiar VL PNe with cool CSs. In the case of
IRE, the model calculations of \cite{SS1999} show that, unlike T$_d$
discussed above, this parameter is less dependent on the evolutionary state
of the PN. However, it depends strongly on the {\em{m}}$_d$/{\em{m}}$_g$ of
the nebula as it is expected to be greater in objects with larger dust
content. This is confirmed by the locations of \DCacr\ and VL PNe in
\figref{dg}, which indicate higher than average ratios of the dust-to-gas
mass in these PNe. On the other hand, the \DCcr\ objects characterized by
{\em{m}}$_d$/{\em{m}}$_g$ at the normal level (just like [WR] PNe, or even
like WEL PNe with {\em{m}}$_d$/{\em{m}}$_g$ below normal) are located in
\figref{ire_Td} to the left of the dotted line with their IRE parameter
being lower than average.

By analyzing the IRE parameter, we find the most puzzling case to be the
\OCacr\ objects. The H\,1-32 nebula with its relatively large
{\em{m}}$_d$/{\em{m}}$_g$ is also characterized by a large IRE in accordance
with what we have just discussed. However, the other \OCacr\ PNe with their
(at most) moderate {\em{m}}$_d$/{\em{m}}$_g$ should have much lower IRE
values. This is not the case. The most extreme example is M\,2-23, which has
the lowest {\em{m}}$_d$/{\em{m}}$_g$ ratio yet its IRE value is among the
largest. By analyzing the model calculations of \cite{SS1999}, it seems that
for \OCacr\ the factor supporting larger IRE values may be their high
density, e.g., due to their slower expansion. According to the models
\citep[Fig.\,13a and 13c of][]{SS1999} IRE settles at an approximately
constant level during an early phase of nebular evolution and becomes quite
insensitive to other parameters. But if the nebular expansion is slow, as in
the case of \OCacr, the time before it occurs is extended and IRE can remain
large for a longer time.

Finally, we should mention that the large IRE value of [WR]-type SwSt\,1
nebula makes it again more similar to \OCacr\ or \DCacr\ than to bulge [WR]
PNe in accordance with the large {\em{m}}$_d$/{\em{m}}$_g$ ratio of this
object that can be inferred from \figref{dg}.

\begin{table*}
\begin{minipage}[t]{1.95\columnwidth}
\caption{Abundances of oxygen and nitrogen adopted from calculations by \cite{Gorny2009}
and compared to earlier published results.}
\centering
\renewcommand{\footnoterule}{}
\begin{tabular}{
               l
               c c @{\hspace{1.40cm}} l} 
\hline
\hline
            & \multicolumn{2}{c}{this work~~~~~~~~~~} & ~~~~~~~~~~~~~~literature \\
            & log O/H+12  & log N/O &  (log O/H + 12; log N/O) \\
\hline
  H 1-16    &  8.83       & -1.06   &  (9.09; -1.32)$^{Sa}$ \\
  Hb 6      &  8.66       & -0.18   &  (8.71; -0.06)$^{AK}$ \\
  M 2-10\footnote{The values adopted are the averages of original calculations
  (see Table\,4) by \cite{Gorny2009} and from \cite{Stasinska1998}.}
            &  8.67       & -0.28   &  (8.78; --)$^G$~~ (8.86; -0.49)$^R$~~ (9.00; -0.55)$^K$~~ (8.90; -0.59)$^{St}$ \\
  H 1-50    &  8.66       & -0.37   &  (8.65; -0.51)$^E$ \\
\vspace{0.1cm}    
  H 1-20    &  8.61       & -0.10   &  (8.97; -0.56)$^{E}$~~ (8.45; --)$^G$~~ (8.97; -0.23)$^R$~~ (9.30; -0.67)$^{Sa}$~~ (8.73; -0.44)$^K$ \\
  Th 3-4    &  8.22       & -0.11   &  \\
  M 3-38    &  8.39       &  0.19   &  (8.37; 0.32)$^R$~~ (8.79; -0.61)$^{Sa}$~~ (8.54; -0.22)$^K$~~ (8.37; --)$^{St}$ \\
  M 3-8     &  8.38       & -0.26   &  (8.59; -0.42)$^E$~~ (9.02; -0.27)$^R$~~ (8.12; -0.27)$^C$~~ (8.15; -0.33)$^{St}$ \\
\vspace{0.1cm}
  H 1-40    &  8.19       & -0.25   &  (9.09; -0.04)$^E$~~ (8.23; -0.45)$^{HS}$~~ (8.70; -0.62)$^R$~~ (8.62; 0.02)$^{St}$ \\
  M 2-23    &  8.35       & -0.99   &  (8.48; -1.44)$^{E}$~~ (8.67; -1.00)$^{HS}$~~ (8.21; -0.13)$^{St}$~~ (8.22; -0.82)$^R$~~ (8.48; 0.20)$^K$ \\
  He 2-260  &  8.00       & -0.76   &  \\
  H 1-32    &  8.59       & -1.21   &  (8.55; -0.23)$^{St}$~~ (8.12; -0.53)$^R$ \\
  H 1-35    &  8.65       & -1.19   &  (8.31; -0.92)$^R$~~ (8.27; -0.89)$^K$ \\
\hline
\hline
\end{tabular}
\end{minipage}

{\em References}: AK -- \cite{AllerKeyes1987}, 
            C -- \cite{Costa1996},
            E -- \cite{Exter2004}, 
            G -- \cite{Gutenkunst2008}, 
            R -- \cite{Ratag1997}\\
            K -- \cite{Koeppen1991},
            Sa -- \cite{Samland1992},  
            St -- \cite{Stasinska1998},
            HS -- \cite{HoofSteene1999}.\\
{\em Note}: Results by \cite{Cuisinier2000}, \cite{Escudero2004}, \cite{Gorny2004} and \cite{WangLiu2007} are not
presented as the line measurements from these papers were used by \cite{Gorny2009}. 
\end{table*}

\subsection{Chemical composition}

We present the chemical composition of the analyzed PNe derived from optical
spectra with the classical empirical method. The values were taken from
\cite{Gorny2009} where the applied method have been described in detail. One
important difference of our study from analysis of \cite{Gorny2009} is that
we do not limit our discussion of \DCcr, \DCacr, and OCacr\ objects to
parameters with errors smaller than 0.3 dex as in the case of some PNe their
spectra do not allow for that quality.

The chemical elements in PNe are frequently divided into two groups. For one
group of abundances it can be safely considered that they remain mostly
unchanged during the previous evolution of the CS and therefore
the values found in PNe represent the primordial abundances of the matter
the progenitor star was born from. An example of such an element is oxygen,
which is regarded as being mostly undisturbed since the object was born, at
least in the case of PNe in metal-rich environments such as the Galactic
bulge \citep[see e.g.,][for detailed discussion]{Chiappini2009}. The
abundances of other elements are however expected to be changed as the
result of various physical processes. Their abundance ratios are modified by
nuclear reactions and mixing that can bring some freshly synthesized matter
to the stellar surface in so-called dredge-up processes. An example of such
an element is nitrogen.

\begin{figure}
\resizebox{\hsize}{!}{\includegraphics{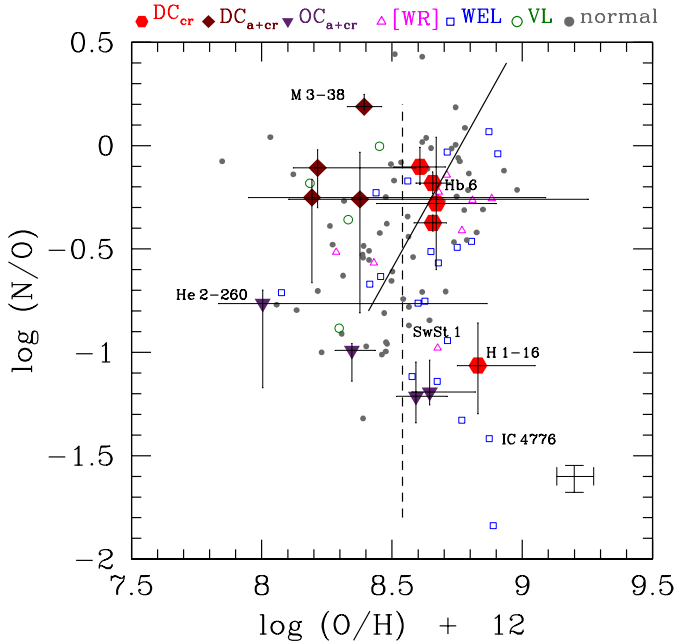}}
\caption[]{ 
  The nebular abundance ratios N/O versus O/H for the different
  groups of GBPNe with meaning of
  the symbols the same as in \figref{l_vlsr}. The solid line marks the relation between O/H 
  and N/O abundances (for PNe with log\,N/O$>$-0.8). The dashed vertical line marks 
  median O/H of \normal\ PNe. 
}
\label{oh_no}
\end{figure}

In \figref{oh_no}, we plot the abundance ratios O/H versus N/O for the
different types of GBPNe. These data can also be found in Table\,3, where
there are compared to earlier results from the literature. As can be noted,
the \DCcr\ objects (except H\,1-16) have locations in this plane that are
compatible with the majority of bulge PNe, including \normal\ ones and those
with [WR] or WEL type CSs. The locations of \DCacr\ are different as they
show a substantial underabundance of oxygen related to both hydrogen and
nitrogen. The \OCacr\ PNe form another extreme with N/O showing the lowest
values among GBPNe, while O/H seems normal. Interestingly, the unusual
objects IC\,4776 and SwSt\,1 are located in the same region of the plot as
\OCacr\ PNe.\footnote{
 We computed chemical abundances of SwSt\,1 and IC\,4776 using line
 intensities from \citet{Pena2001} and \citet{Exter2004}}

Before considering the possible reasons of this behavior, one must first
discuss whether the obtained oxygen abundances are reliable. As
directly indicated in \figref{oh_no} by the errorbars, the formal errors of
derived abundances are sometimes very large, especially for the \DCacr\ PNe.
They were calculated by propagating the possible observational errors
of measured spectral lines into computed parameters using the Monte Carlo
method \citep[for details see][]{Gorny2009}. The typical errors for the other PNe
presented in \figref{oh_no} are indicated with the errorbar cross in the
bottom-right corner of the plot.  

Despite the large individual errors, we observe that the \DCacr\ PNe clearly
have oxygen underabundances of more similar value than expected, because the
random errors should produce a greater scatter in the distributions than
observed. In \figref{oh_no}, we mark with a solid thick line the relation
between O/H and N/O abundances for GBPNe with log\,N/O$>$-0.8 and good
quality data (rejecting only some clear outliers). Assuming that this holds
for all PNe and that observed deviations from it are caused only by
individual errors the probability of finding all four \DCacr\ at their
present locations in \figref{oh_no} can be evaluated as being smaller than
1\%. This hypothesis can therefore be safely rejected meaning that the O/H
vs. N/O relation for \DCacr\ is truly significantly different from that of
other GBPNe. Also using the Kolmogorov-Smirnoff 2D nonparametric test, the
hypothesis that the difference between \DCacr\ and
\normal\ PNe in the \figref{oh_no} is statistically meaningful is confirmed
at about 99\% confidence level. That for the \DCacr\ object with the
smallest errors (M\,3-38) we infer the same effect of oxygen
underabundance\footnote{
 In fact, for M\,3-38 we have two independent, high quality spectra that provide
 very similar results.} 
is also very important.

\begin{figure}
\resizebox{\hsize}{!}{\includegraphics{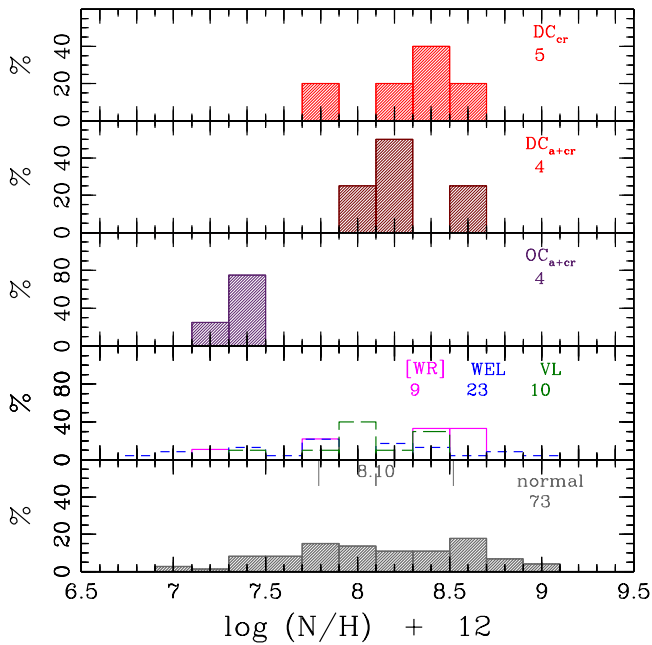}}
\caption[]{
  Distributions of N/H abundance ratio for the different groups of Galactic
  bulge PNe. The same notations apply as in \figref{hist_Ne}.
}
\label{hist_nh}
\end{figure}

\begin{figure*}
\resizebox{.49\hsize}{!}{\includegraphics{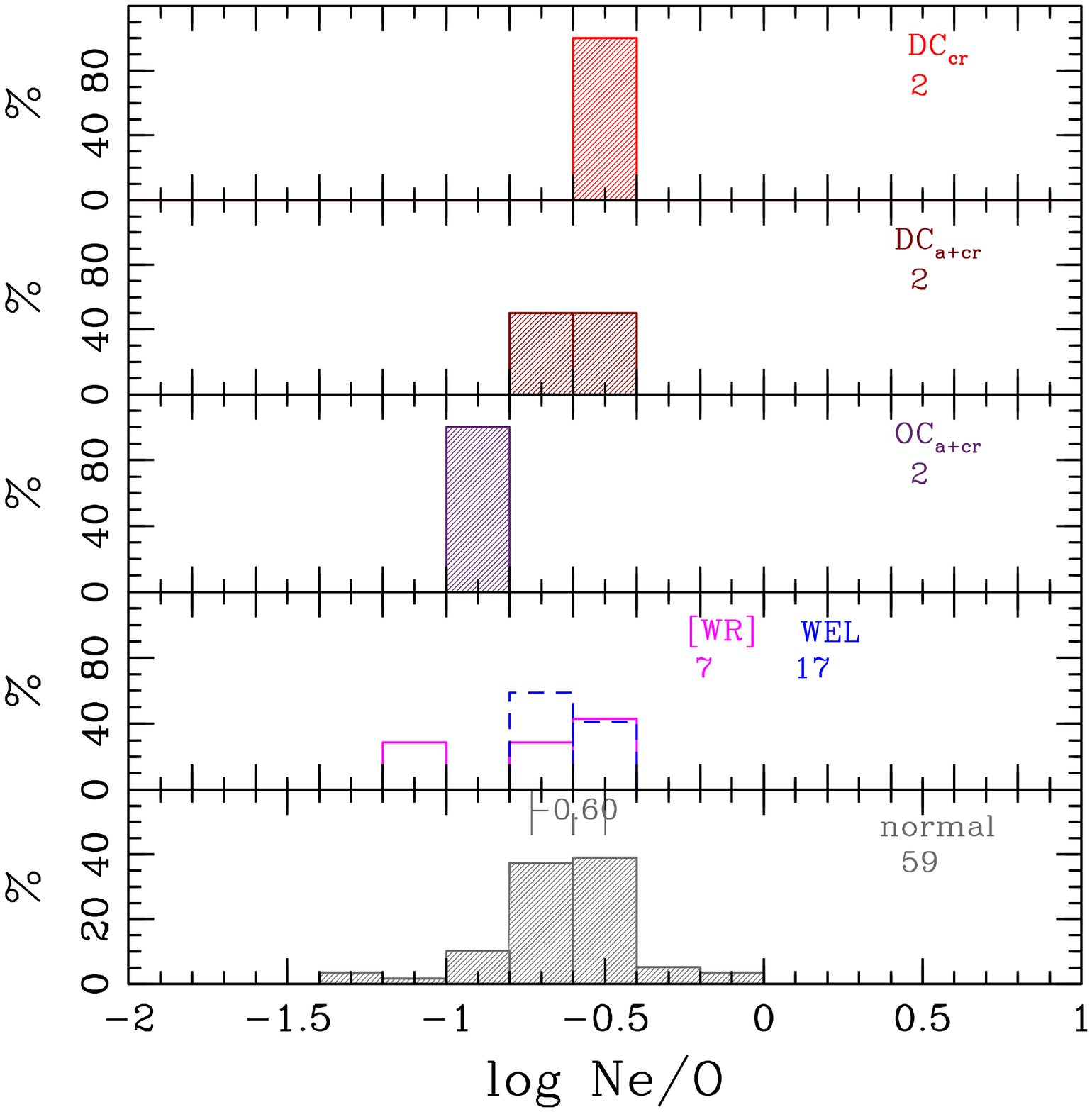}}
\resizebox{.49\hsize}{!}{\includegraphics{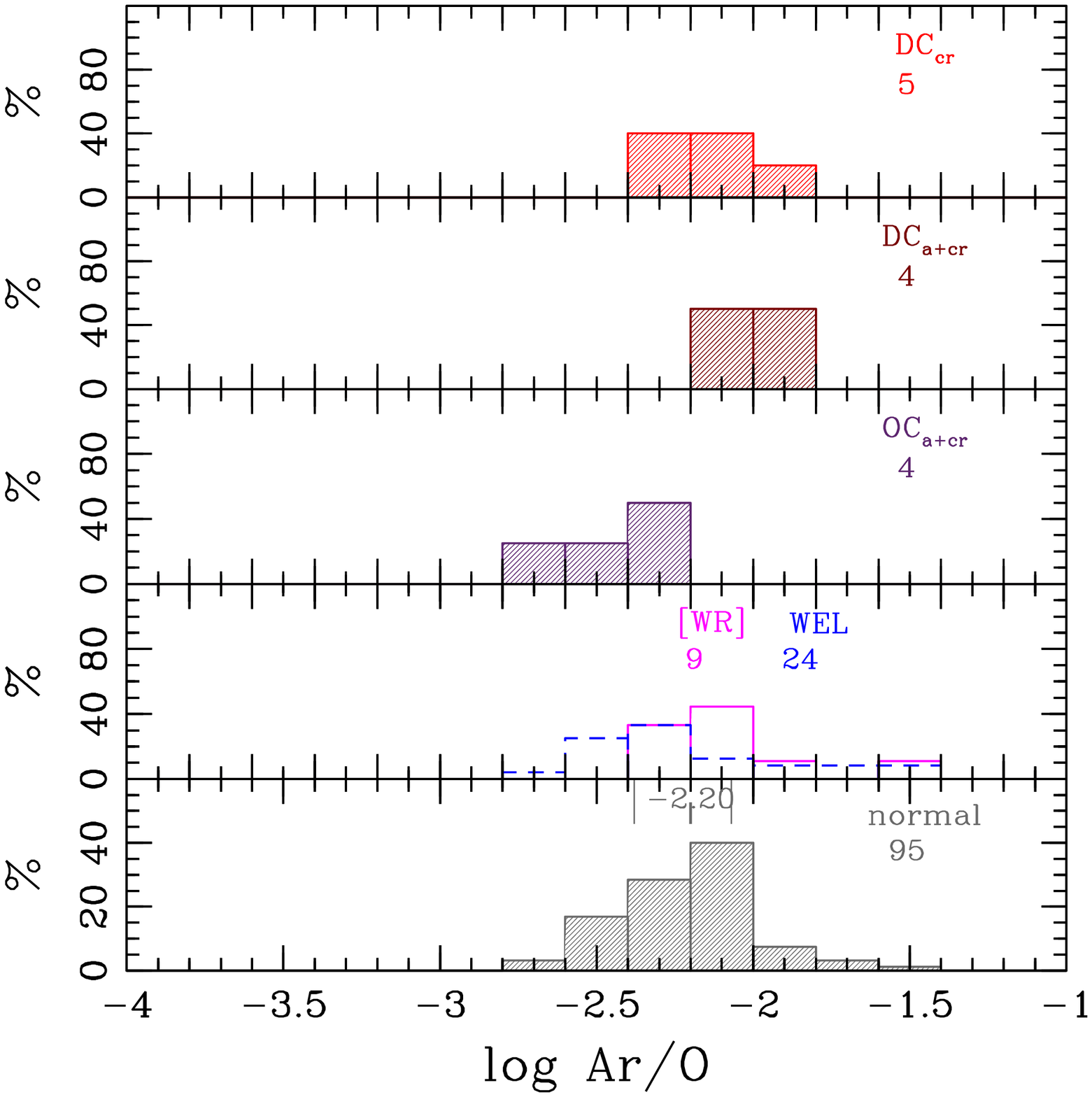}}

\resizebox{.49\hsize}{!}{\includegraphics{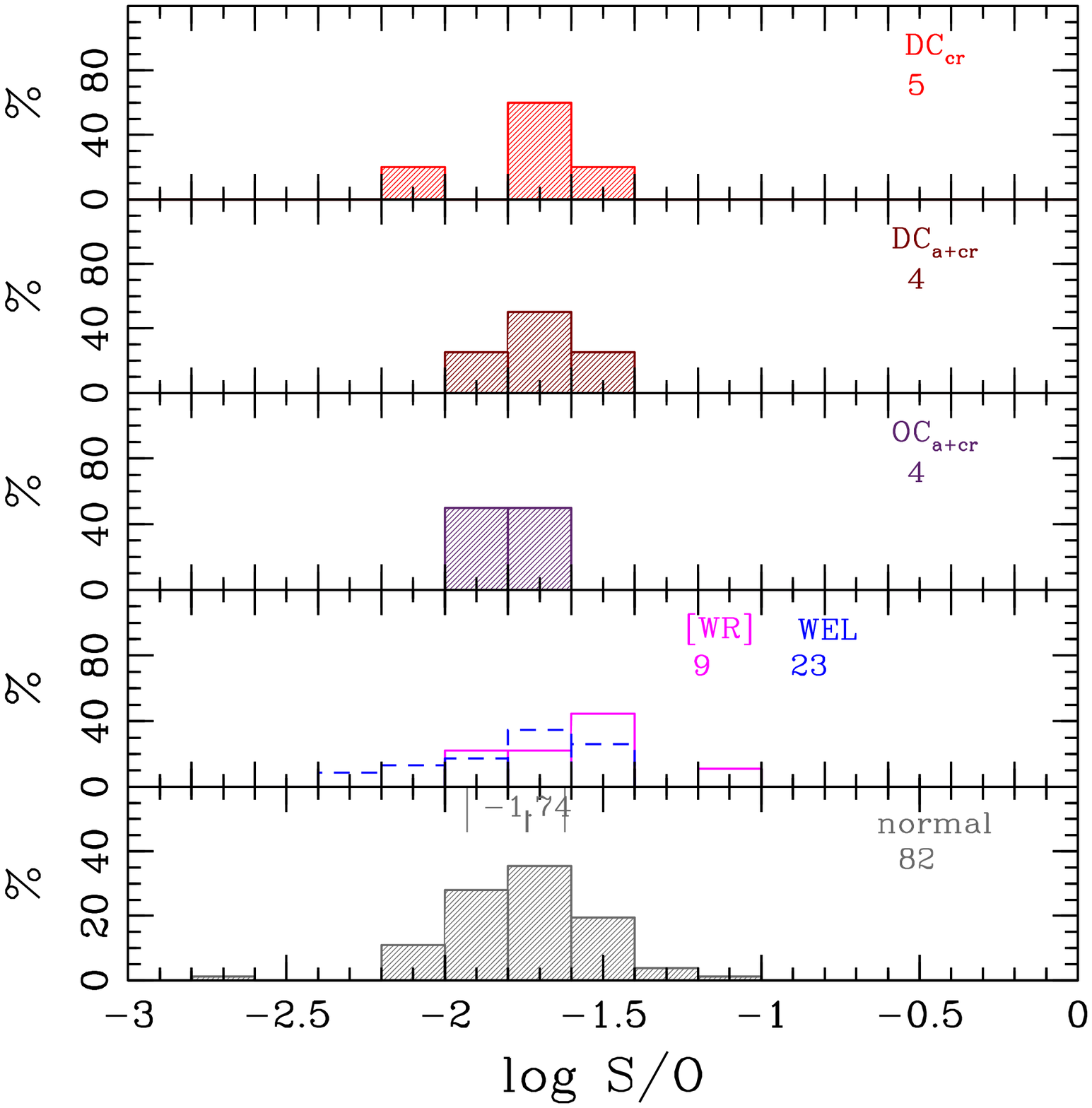}}
\resizebox{.49\hsize}{!}{\includegraphics{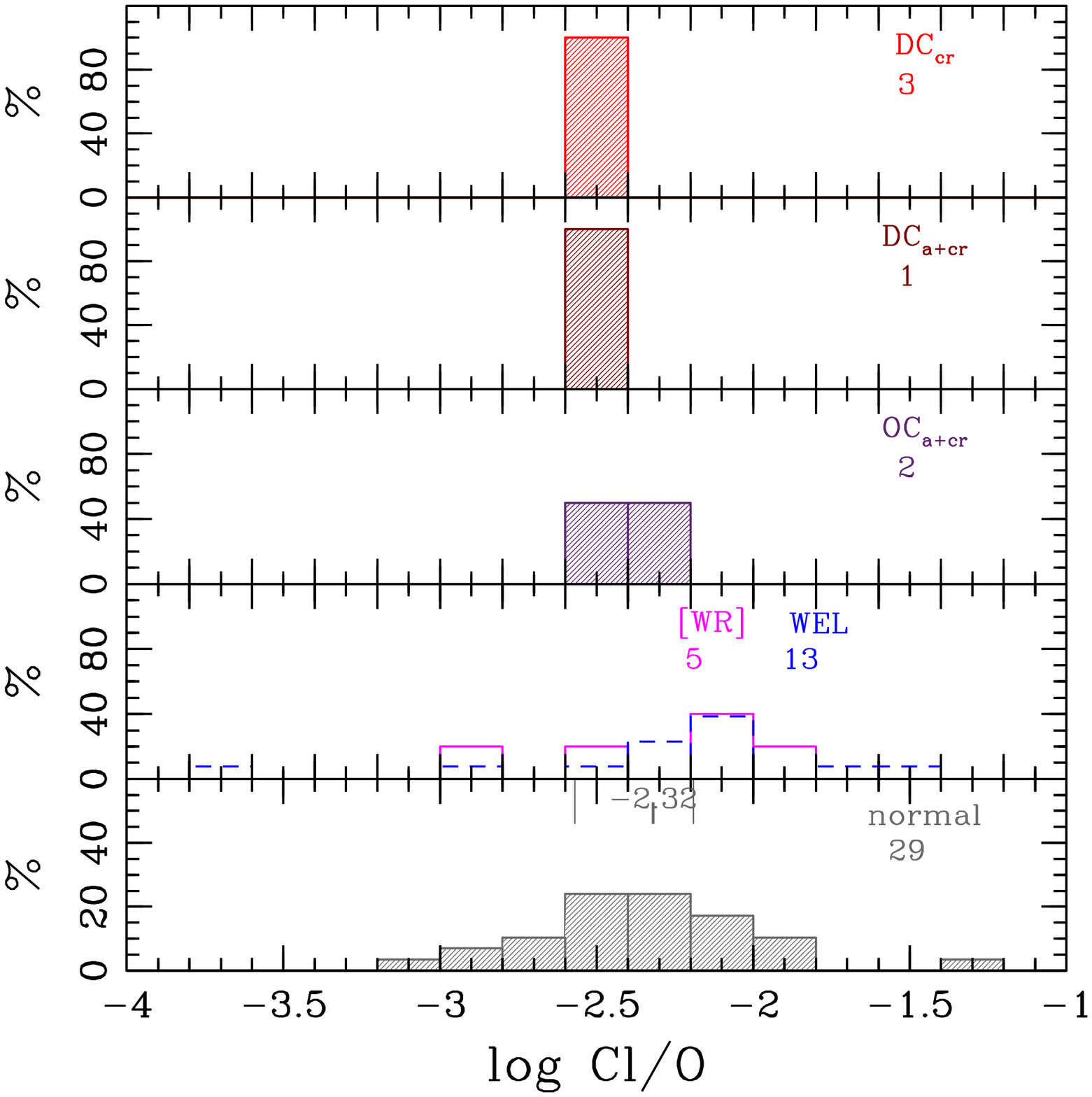}}
\caption[]{
  Distributions of abundance ratios for different groups of Galactic bulge
  PNe.  For the \normal\ PNe, the median value along with the 25 and
  75 percentiles are marked with three short vertical lines above the
  histogram. Total number of objects included are given below sample names. 
}
\label{hist_Xo}
\end{figure*}

In Cols.\,3 to 9 of Table~4. are list the abundances of nitrogen
and other elements for PNe investigated here and the median values for
\normal, [WR], and WEL GBPNe.  In \figref{hist_nh}, we present the
histograms of the N/H ratio for the different types of GBPNe. Comparing the
\DCacr\ PNe with \normal\ PNe, one can see no enhancement of
nitrogen as the median values are almost the same.  In the case of neon,
sulphur, and possibly chlorine there seems to be no depletion of these
elements relative to oxygen (see \figref{hist_Xo} available online). This
means that whatever the reason for the lower oxygen abundance in
\DCacr\ PNe a similar decrease in the abundances of these three elements
follows. Only in the case of argon is there possibly a difference at the
level of 0.1dex in the median abundance ratio of Ar/O between \DCacr\ and
\normal\ PNe. 

By analyzing the histograms of \DCcr\ objects in Figs.\,\ref{hist_nh} and
\ref{hist_Xo} there is no indication of any important difference
in chemical abundances between this group and \normal\ PNe.

The situation of \OCacr\ PNe is different. As can be deduced from the
histogram in \figref{hist_nh}, their underabundance of N/O is caused
directly by the considerably lower number of nitrogen atoms. By analyzing
the data for other elements in \figref{hist_Xo}, we also find that the
abundances of neon and argon seem lower than the average levels in
\normal\ PNe. On the other hand, no difference is noticeable for the
abundances of chlorine and sulphur.

\section{Discussion}

There has been a long accepted opinion that post-AGB objects can belong to
only one of the two groups: either those characterized by an oxygen-rich
environment or those surrounded by carbon-dominated matter. However, soon
after the first observations of Galactic PNe by ISO satellite were released
it become clear that objects exist, that is [WR] PNe with late-type CSs,
that simultaneously contain both carbon-based and oxygen-based dust. A
number of hypotheses have been invoked to explain this phenomenon
\citep[see the review in][]{P-C2009}. However, the dual-dust chemistry
phenomenon remained to be viewed as a rare event among PNe that happens to
only some objects. Furthermore, these objects were already acknowledged to
be very peculiar because of the hydrogen-deficient composition of the CSs
and strong WR-type stellar winds.

The Spitzer observations of GBPNe \citep{Gutenkunst2008,P-C2009} yielded a
number of discoveries. The first surprise was a higher than expected
percentage of PNe with dual-dust chemistry, which as we discussed in the
Introduction might have been partly caused by the lower sensitivity of ISO.
\cite{Gutenkunst2008} argued that dual-dust chemistry may be related
to the binary evolution and the existence of circumbinary disks. They
attributed the observed rate to a higher proportion of [WR] PNe in the
Galactic bulge, which has not been actually proven \citep{Gorny2009},
the \cite{Gutenkunst2008} sample containing only one such object.

Analyzing their sample of GBPNe \cite{P-C2009} noted some other important
results. The first was that dual-dust chemistry phenomenon is present not
only in PNe surrounding late type [WC] CSs but in all [WR] PNe and also in
all other PNe with emission-line CSs (WEL, VL) as well as many PNe without
emission-line CSs. \cite{P-C2009} also found that the percentage of PNe with
C-rich material is confusingly high in the Galactic bulge since there is a
well known deficit of C-rich AGB stars in that environment.
We note that all GBPNe that contain PAHs are dual-dust chemistry sources.

\cite{P-C2009} noted that the population of O-rich AGB stars in
the Galactic bulge, with the exception of obscured OH/IR stars
\citep{Vanhollebeke2007}, do not show any indication of  crystalline
silicates in their spectra. Thus the high detection rate of dual-dust
chemistry found in PNe cannot be explained by long-lived O-rich (primordial
or circumbinary) disks. That most of GBPNe cannot originate in the binary
systems is indicated also by the results of \citet{Miszalski2008}. They
reported that among 300 analyzed GBPNe only 21 (about 12-21\%) show the
signatures of periodic variability in the OGLE-III data, which may be
interpreted as being caused by close binarity.

The low-mass O-rich AGB stars in the Galactic bulge cannot bring enough
carbon into the envelope to produce C-rich AGB stars since in the
high-metallicity environment the efficiency of their third dredge-up is very
limited.  Therefore, \citet{P-C2009} proposed that the most plausible
scenario for creating C-rich AGB stars seems to be a {\em final} thermal
pulse on the AGB (or just after), which would produce an enhanced mass loss,
capable of removing/mixing (sometimes completely) the remaining H-rich
envelope and exposing the C-rich layers. It would also generate shocks
responsible for silicate crystallization in the ejected circumstellar shell.

For [WR] PNe there are many arguments that the change in the composition of
the CS occurs predominantly on the AGB or soon after
\citep{GornyTylenda2000}. Since dual-dust chemistry in the Galactic disk PNe
had been revealed in [WR]-type objects it was natural to expect that a
successful scenario should explain simultaneously both the unusual
composition of [WR]-type CSs and the dual-dust chemistry of their nebulae
\citep{Gorny2008}. The {\em final} thermal pulse at the end of the AGB
suggested by \cite{P-C2009} as a most plausible theory meets this
expectation. However, it is unclear why dual-dust chemistry is so widely
observed among GBPNe and restricted to only some Galactic disk objects.
\cite{P-C2009} pointed out that metallicity may be a possible explanation
since the metallicities of GBPNe are on average higher than those in the
Milky Way disk.

The properties of GBPNe with emission-line CSs of various types were
investigated by \cite{Gorny2009}. In the present paper, we have therefore
concentrated on those PNe that do not contain emission-line CSs. After
inspecting the Spitzer spectra, we found that they can be divided into
three\footnote{
  There exist in the Galactic bulge also PNe with only
  signs of crystalline silicates but as explained in Sect.\,2 we did not
  have enough data to discuss them as a separate class here.} clearly
separate groups \DCcr, \DCacr, and \OCacr. Objects of the first
group, \DCcr, have simultaneously both carbon-based dust
(PAHs) and oxygen-based dust (crystalline silicates). In the second group,
\DCacr, there are both PAHs and crystalline
silicates, but in addition there are also amorphous silicates. In the last
group, \OCacr, there is only oxygen-rich dust but in both crystalline and
amorphous forms. Our primary discovery is that this classification based on
the dust features in the infrared spectra is reflected in some other, more
general properties of PNe.

The \DCcr\ PNe have infrared spectra that most closely resemble those of PNe
with emission-line CSs. One can note the similar signatures of crystalline
silicates at 23.5, 27.5, and 33.8\,\um\ and of PAHs at 6.2, "7.7", 8.6, and
11.3\,\um. At the same, time the chemical composition of \DCcr\ nebulae
cannot be distinguished from that of the WEL objects and in general from the
majority of GBPNe.  In principle, one could consider whether \DCcr\ are not
an earlier evolutionary phase of WEL PNe when stellar emission lines are not
yet visible but our analysis of their evolutionary state do not give strong
support for such possibility. It cannot however be completely ruled out that
for some \DCcr\ PNe, stellar emission-lines have not yet been discovered
because of the quality of the spectra.

For the possible links between \DCcr\ and [WR] PNe, it has to be
noted that the latter objects are definitely brighter (see Fig.\,4) and
have more massive nebulae (Fig.\,6). Both differences could be explained if
\DCcr\ PNe are less evolutionary advanced. But in that case, it is hard to
accept that relatively strong stellar emission-lines (expected spectral
types should range from [WC11] to [WC7]) could remain unnoticed in the
optical spectra. We can also add that large-scale turbulent motions
are characteristic of PNe around [WR]-type CSs \citep{Gesicki2006}, however no
information about them have so far been reported for any of \DCcr\ PNe.

The objects belonging to \DCacr\ and \OCacr\ groups exhibit far more
pronounced differences from the other GBPNe.  The two groups are linked by
showing evidence of amorphous silicates at 10\um\ and belonging to the
densest PNe in the observed Galactic bulge population.  However, the other
observational results imply that the evolutionary status of \DCacr\ and
\OCacr\ must be completely different.

The high densities of \DCacr\ PNe can be regarded as a sign of the
relatively short time that has passed since they left the AGB but it is more
likely caused by their nebulae being more clumpy. \DCacr\ are also
characterized by the considerable excess in extinction that could be
attributed to some source of internal extinction. Finally, the nebular gas
has a peculiar chemical composition with oxygen being underabundant relative
to hydrogen, nitrogen, and possibly argon but preserving normal levels when
compared to other elements. This seems difficult to interpret in the
framework of the standard chemical evolution of PNe progenitors. If the O/H
in \DCacr\ PNe were to represent the primordial oxygen abundance of the matter
from which they were formed, it would favor older objects born before the
interstellar matter was enriched in metals. In that case however, the CSs of
\DCacr\ PNe should be slowly evolving low mass objects and the N/O ratio
should not be increased since the effective dredge-up of nitrogen occurs in
higher mass AGB stars. In contrast, our analysis of the evolutionary status, indicate
that \DCacr\ have intermediate and sometimes clearly higher mass CSs.

The low O/H abundance accompanied by a higher than normal N/O ratio is
possible if the ON cycle of nuclear reactions was active in the progenitor
stars. In that case the O/H ratio no longer reflects the primordial oxygen
content. The ON cycle seems to work preferentially within more massive
stars in low-metallicity environments. The examples are some PNe from
LMC that show a clear anti-correlation between O/H and N/O ratios
(\cite{LeisyDennefeld2006}, see also \cite{Chiappini2009} and the discussion
therein). On the other hand, the ON cycle should have no effect on other
elements. For the \DCacr\ PNe discussed here, this is not the case as
the abundances of neon, sulphur and chlorine closely follow the depletion of
oxygen.

Finally, in \DCacr\ PNe some depletion of oxygen could be possible because
it is being trapped in dust grains. The percentage of oxygen removed in this
way from the gas to the dust may be metallicity dependent and is limited by
the amount of silicon available. More could be depleted by means of ice
growth but this would give rise to recognizable features in the infrared
spectra. The depletion may simultaneously concern not only oxygen but also
some other elements such as sulphur. However, it should have no effect on
noble gas such as neon. As the Ne/O ratios of \DCacr\ nebulae are not
enhanced, this hypothesis should also be ruled out.

The \OCacr\ PNe are characterized by very small diameters and the highest
densities but at the same time expand more slowly than other GBPNe.  For
this reason, their present evolutionary state indicates they should be related
to intermediate mass CSs ($\approx$0.60\Msun). However, this is inconsistent
with the very low metallicity of their surrounding nebulae as seen
in the low N/O ratio derived for these objects as well as of neon and argon.
In contrast, low metallicity argues for the low mass CSs originating
in lower mass progenitors that were created before the Galactic bulge was
effectively enriched in metals\footnote{
  In that case, the O/H ratio may not represent the primordial
  composition, as is the case of PNe in the LMC - see \cite{Chiappini2009}}.

The evolution of a low mass CS can be accelerated by high mass-loss.
\cite{Kudritzki1997} detected very strong winds from \OCacr\ object H\,1-35
(the final numerical value has not been given). On the other hand, He\,2-260
has normal, low mass-loss at the level of 0.45$\times$10$^{-7}$\,\Msun\//yr
\citep{Hultzsch2007}. In both cases the chemical composition of the CS is
normal, i.e., they are not H-deficient.

The possible solution to the puzzle of the relatively fast evolution in the
nebular phase and the low metallicity of \OCacr\ could be their origin not
from single stars but from a binary system including a lower mass star. This
star would not normally be observed in the PNe phase because its post-AGB
evolution is too slow to ionize the ejected gas before it disperses.
However, as a result of the mass transfer from the companion at earlier
evolutionary phases of the binary system (or even merging in extreme cases),
the future CS could increase its mass and evolve more rapidly during its
post-AGB phase. The PN would become visible, however, the nebular abundances
of some elements possibly being characteristic of a lower mass (older)
progenitor.

The evolution of the AGB star in the binary system can have other
consequences. As argued by \cite{DeMarco2009} for nebulae created from
common-envelope binaries, the abundances of e.g. nitrogen and carbon should
be statistically lower that in "normal" PNe. This is because the interaction
with the companion will cause the AGB star to terminate this phase of
evolution earlier than of the single star. As a result, no effective
dredge-up will take place associated with the thermal pulses at the tip of
AGB. There will also be no {\em final} thermal pulse. The nebula M\,2-23
was investigated by \cite{Miszalski2008} but no photometric variations were
found. Nevertheless, the low N/O ratio is the characteristic feature of
\OCacr\ PNe. We note also that the \OCacr\ are the only GBPNe that clearly do
not have C-rich dust. At the same time, not all the surrounding silicate
grains have been crystallized. All these results imply that the formation of
\OCacr\ PNe differs from that of the rest of the Galactic bulge
population.

This leads us to the main conclusion of our work: there is clearly no unique
road to the formation of PNe even in a uniform environment such as the
Galactic bulge. This is noteworthy since the stars in the Galactic bulge
that we are now able to observe in their PNe phase are expected to originate
mostly from a single episode of star formation. Obviously, there are PNe in
the Galactic bulge with very different CSs and different chemical
compositions. They are also characterized by different properties of the
dust as seen in \DCcr, \DCacr, and \OCacr\ groups investigated in this
paper. Nonetheless, the simultaneous presence of PAHs and crystalline
silicates dominates in the GBPNe. Therefore, the scenario of {\em final}
thermal pulse at the end of the AGB that changes both the stellar
composition to C-rich and at the same time allows the crystallization of
existing O-rich grains remains the most plausible possibility for the
majority of GBPNe. However, it is not always effective or in different ways
for different stars.

\section{Conclusions}

We have investigated PNe without emission-line central stars located towards
the Galactic bulge that have peculiar infrared spectra acquired by
Spitzer/IRS. Among these objects, we have found three separate groups
divided according to their composition of dust grains:

\begin{itemize}

\item{\DCcr\ -- dual-dust chemistry PNe with simultaneous presence of both
      carbon-based dust (PAHs) and oxygen-based dust (crystalline
      silicates);}

\item{\DCacr\ -- dual-dust chemistry PNe with simultaneous existence of PAHs and
     crystalline silicates as well as amorphous silicates;}

\item{\OCacr\ -- PNe characterized by oxygen dust chemistry with only
      oxygen-rich grains in both crystalline and amorphous forms.}

\end{itemize}

\noindent
We have analyzed a wide range of different properties of these PNe. Our main
results are:

\begin{itemize}

\item{We confirm that dual-dust chemistry is a common phenomenon of PNe
      in the Galactic bulge and can occur in objects not related to
      emission-line central stars.}

\item{The Properties of \DCcr\ PNe do not distinguish them clearly from the
      majority of other PNe in the Galactic bulge. They have intermediate or
      higher-mass central stars. Their infrared spectra closely resemble
      those of PNe with emission-line nuclei. Some \DCcr\ may be
      evolutionary related to the latter objects or may have undiscovered
      emission-line central stars.}

\item{\DCacr\ objects belong to the densest PNe in the Galactic bulge. Their
      derived {\em{m}}$_d$/{\em{m}}$_g$ mass ratios and infrared excesses
      IRE are higher than average. There is a possibility of extensive
      internal extinction. \DCacr\ PNe have intermediate and higher-mass
      central stars.  The chemical composition of nebular gas is peculiar as
      oxygen seems underabundant relative to hydrogen and nitrogen but not
      to other elements (except possibly argon). This composition of \DCacr\
      PNe cannot be explained in the standard picture of AGB star
      chemical evolution.}

\item{The \OCacr\ PNe are the only analyzed PNe not showing dual-dust chemistry.
      They have hottest dust temperature T$_d$ and highest infrared excess IRE.
      \OCacr\ have also very small diameters and are among the densest PNe in
      the Galactic bulge. However, their expansion velocities are smaller
      than average and therefore their evolutionary status indicates that
      \OCacr\ can have intermediate-mass central stars. In contrast,
      the surrounding nebulae show low metallicity with an underabundance of
      nitrogen, neon, and argon. The domination of oxygen-based dust
      indicates in addition a low abundance of carbon. We argue that their
      properties are in qualitative agreement with scenarios of PNe
      formation not from single AGB stars but from binary systems.}

\end{itemize}

\begin{acknowledgements}
 We acknowledge support from the Faculty of the European Space Astronomy
 Centre (ESAC) and from the Comunidad de Madrid PRICIT project
 S-0505/ESP-0237 (ASTROCAM). R.Sz. and S.K.G. acknowledge support from
 grant N203 393334 of the Science and High Education Ministry of Poland. 
 D.A.G.H. acknowledges support for this work provided by the
 Spanish Ministry of Science and Innovation (MICINN) under the 2008 Juan 
 de La Cierva Program and under grant AYA-2007-64748.
\end{acknowledgements}

\bibliographystyle{aa}
\bibliography{13010}

\Online

\begin{table*}
\caption{Chemical abundances for analyzed PNe and median values for bulge [WR], WEL, and \normal\ PNe samples \citep[from][]{Gorny2009}.}
\begin{tabular}{
               l
               l @{\hspace{0.20cm}}
               r r c c c c c}
\hline
\hline
 PN G       & name       
                         
                         & He/H     &   N/H    &   O/H    &  Ne/H    &   S/H    &  Ar/H    &  Cl/H    \\
\hline
 000.1+04.3 &  H 1-16    &  11.05    & 5.83E-05 & 6.76E-04 &      -   & 6.08E-06 & 2.94E-06 &      -   \\
 007.2+01.8 &  Hb 6      &  11.09    & 2.99E-04 & 4.53E-04 & 1.47E-04 & 9.84E-06 & 3.72E-06 & 1.20E-06 \\
 354.2+04.3 &  M 2-10    &  11.15    & 2.99E-04 & 2.73E-04 &      -   & 7.93E-06 & 3.58E-06 & 7.69E-07 \\
 358.7-05.2 &  H 1-50    &  11.07    & 1.92E-04 & 4.54E-04 & 1.31E-04 & 7.36E-06 & 2.35E-06 & 1.63E-06 \\
 358.9+03.2 &  H 1-20    &  11.16    & 3.20E-04 & 4.05E-04 &      -   & 7.51E-06 & 3.98E-06 &      -   \\
 354.5+03.3 &  Th 3-4    &  11.04    & 1.28E-04 & 1.64E-04 & 3.38E-05 & 2.98E-06 & 1.57E-06 &      -   \\
 356.9+04.4 &  M 3-38    &  11.11    & 3.81E-04 & 2.47E-04 & 6.36E-05 & 7.50E-06 & 2.82E-06 & 8.25E-07 \\
 358.2+04.2 &  M 3-8     &  11.13    & 1.31E-04 & 2.38E-04 &      -   & 3.47E-06 & 2.53E-06 &      -   \\
 359.7-02.6 &  H 1-40    &  11.05    & 8.74E-05 & 1.56E-04 &      -   & 3.01E-06 & 1.17E-06 &      -   \\
 002.2-02.7 &  M 2-23    &  11.05    & 2.27E-05 & 2.22E-04 & 3.46E-05 & 3.56E-06 & 8.63E-07 & 1.05E-06 \\
 008.2+06.8 &  He 2-260  &   9.96    & 1.74E-05 & 1.01E-04 &      -   & 1.67E-06 & 2.14E-07 &      -   \\
 355.6-02.7 &  H 1-32    &  11.05    & 2.39E-05 & 3.91E-04 &      -   & 4.36E-06 & 1.79E-06 &      -   \\
 355.7-03.5 &  H 1-35    &  11.06    & 2.84E-05 & 4.42E-04 & 5.80E-05 & 6.28E-06 & 1.93E-06 & 1.31E-06 \\
\hline
 [WR]       &            &  11.12    & 2.63E-04 & 4.79E-04 & 5.89E-05 & 1.26E-05 & 4.07E-06 & 2.34E-06 \\
 WEL        &            &  11.08    & 7.41E-05 & 4.47E-04 & 1.10E-04 & 4.90E-06 & 2.04E-06 & 3.02E-06 \\
 \normal\   &            &  11.14    & 1.26E-04 & 3.47E-04 & 7.76E-05 & 5.89E-06 & 2.09E-06 & 1.51E-06 \\
\hline
\hline
\end{tabular} 
\end{table*}

\clearpage

\end{document}